\documentclass{article}

\PassOptionsToPackage{numbers, compress, sort}{natbib}

\usepackage[preprint]{neurips}

\usepackage[utf8]{inputenc} 
\usepackage[T1]{fontenc}    
\usepackage{url}            
\usepackage{booktabs}       
\usepackage{amsfonts}       
\usepackage{nicefrac}       
\usepackage{microtype}      
\usepackage{xcolor}         

\usepackage{amsmath}
\usepackage{booktabs}
\usepackage{multirow} 
\usepackage{pifont}
\usepackage{graphicx} 
\usepackage{textcomp}
\usepackage{url}
\usepackage{balance}
\usepackage{makecell}
\usepackage{adjustbox}
\usepackage{forest}
\usepackage{subcaption}
\usepackage{wrapfig}
\usepackage{tablefootnote}
\usepackage{threeparttable}
\usepackage{xspace}

\usepackage{colortbl}
\usepackage{color}
\usepackage{xcolor}
\usepackage{makecell}
\usepackage{enumitem}
\usepackage{listings}

\definecolor{lightTeal}{HTML}{add2ca}
\definecolor{softPurple}{HTML}{8c89ba}
\definecolor{paleGreen}{HTML}{e6f0db}
\definecolor{mutedOrange}{HTML}{dfa660}
\definecolor{limeGreen}{HTML}{97c04b}
\definecolor{lightBlue}{HTML}{dbeef3}
\definecolor{softRed}{HTML}{d77470}
\definecolor{dustyBlue}{HTML}{93b2d9}
\definecolor{tabletitle}{HTML}{E5DBF8} 

\definecolor{checkgreen}{HTML}{009900}
\definecolor{crossred}{HTML}{CC0000}

\usepackage{mdframed}
\newcounter{rq} 
\newcommand{\answerRQ}[1]{\refstepcounter{rq}
\vspace{1mm}
\begin{mdframed}[linecolor=gray,roundcorner=12pt,backgroundcolor=gray!15,linewidth=3pt,innerleftmargin=2pt, 
leftmargin=0cm, rightmargin=0cm, topline=false, bottomline=false, rightline=false]
\textbf{RQ\arabic{rq} Summary:} #1
\end{mdframed}
}

\newcommand{\appname}{\textsc{ExLabel}\xspace}
\newcommand{\appnamebold}{\textsc{\textbf{ExLabel}}\xspace}
\newcommand{\bench}{\textsc{SWE-TQ}\xspace}
\newcommand{\benchpy}{\textsc{SWE-TQ-Python}\xspace}
\newcommand{\benchml}{\textsc{SWE-TQ-Multilingual}\xspace}
\newcommand{\benchpyml}{\textsc{SWE-TQ-Python\slash Multilingual}\xspace}

\newcommand{\upp}[1]{\textcolor{crossred}{\,\textuparrow #1\%}}
\newcommand{\downp}[1]{\textcolor{checkgreen}{\,\textdownarrow #1\%}}

\newcommand{\fixed}[1]{}

\AtBeginDocument{%
  }

\usepackage{microtype}
\usepackage[hidelinks]{hyperref}       

\title{Well-Specified and Well-Tested? Assessing SWE-bench-Style Datasets with a Self-Evolving Agent}

\author{%
  Zhonghao Jiang$^{1,2,*}$ \quad Xuyang Zheng$^{1,2,*}$ \quad Yanfu Yan$^{1,3,\dagger}$ \\
  \bfseries Zhongxin Liu$^{1,2}$ \quad Lingfeng Bao$^{1,2}$ \quad Shuiguang Deng$^{1}$ \\
  $^{1}$College of Computer Science and Technology, Zhejiang University \\
  $^{2}$State Key Laboratory of Blockchain and Data Security, Zhejiang University \\
  $^{3}$ Hangzhou High-Tech Zone (Binjiang) 
  Institute of Blockchain and Data Security \\
  \texttt{\{zhonghao.j,zhengxuyang,yanfu,liu\_zx,lingfengbao,dengsg\}@zju.edu.cn} \\
}

\begin{document}

\maketitle
{\renewcommand{\thefootnote}{*}\footnotetext{Equal contribution.}}
{\renewcommand{\thefootnote}{$\dagger$}\footnotetext{Corresponding author.}}

\begin{abstract}

Large language models (LLMs) have demonstrated strong performance on software engineering (SE) tasks, driving the need for high-quality datasets to support reliable training and evaluation.
SWE-bench-style datasets (SWE datasets) are widely used to evaluate LLM-based coding agents for resolving real-world software issues, where task quality, particularly task clarity and test coverage, directly affects whether instances are solvable and verifiable.
However, quality issues remain in existing SWE datasets, while manual annotation and verification are costly and labor-intensive, motivating recent efforts toward LLM-based annotation.
Existing approaches typically rely on \textit{static} and overly general judgment rubrics, limiting their ability to accommodate diverse and nuanced scenarios and often resulting in high false positive rates.
In contrast, human annotation usually draws on accumulated experience to make context-aware judgments.

Inspired by this, we propose \appname, an LLM-based, self-evolving, experience-guided assistant for task clarity and test coverage assessment in SWE datasets.
\appname introduces an evolving experience bank that accumulates cross-instance knowledge into stable yet \textit{adaptive} judgment criteria and reuses relevant experience to guide the assessment of new instances.
To enable systematic evaluation, we construct \bench, a new benchmark spanning eight programming languages, by cleaning human annotations from SWE-bench and Multi-SWE-bench.
Experiments with three backbone LLMs show that \appname achieves the best F1 across all settings, with relative improvements of 44.2\%/49.9\% and 24.0\%/18.4\% over the best-performing baseline for task clarity and test coverage assessment on the Python/Multilingual subsets, respectively.
Furthermore, a coding model fine-tuned on SWE-rebench data filtered by \appname resolves 66.2\% of the issues on SWE-bench Verified, outperforming the model trained on data filtered by the best-performing baseline by 14.5\%.

\end{abstract}


\section{Introduction}

Developing and evaluating the software engineering (SE) capabilities of large language models (LLMs) has attracted considerable research attention~\cite{Swe-smith, trae2025, jiang2025agentic}.
SE datasets provide a critical foundation for equipping LLMs with SE-related capabilities~\cite{yu2025utboost, spice, wang2025solved}, serving both as benchmarks~\cite{swebenchverified, nocode-bench, multiswebench} for evaluating model performance on SE tasks and as valuable corpora for model training~\cite{Swe-smith, SWE-Synth, Bugpilot}. 
Among these datasets, SWE-bench-style datasets (\textit{i.e.,} SWE datasets)~\cite{SWE-bench, SWE-rebench, r2egym} have become increasingly popular due to their demonstrated effectiveness in training and evaluating LLM-based coding agents for real-world software issue resolution~\cite{trae2025, openhands}. 
Each instance in SWE datasets consists of a natural-language task description, a golden patch, a set of test suites for evaluating the success of the task, and a corresponding sandboxed executable environment.

An important yet not well-explored aspect of SWE datasets is \textit{task quality}, which concerns the solvability and verifiability of instances~\cite{bosu2013data}. 
Poor task quality not only undermines evaluation reliability but can also mislead model training and benchmarking, leading to inflated or unreliable performance estimates. 
Despite its importance, quality issues have been observed in existing SWE datasets~\cite{liveswebench, Swe-mirror, SWE-Lego, Swe-smith}. 
Issues such as insufficient test coverage~\cite{wang2025solved, yu2025utboost} persist even in the manually verified SWE-bench dataset~\cite{swebenchverified}. 
For SWE datasets containing synthetic data~\cite{Swe-smith, Swe-mirror}, such issues may be further exacerbated, as large-scale automated data generation may introduce additional noise and inconsistencies.
Task quality is commonly assessed along two dimensions~\cite{swebenchverified}: \textit{task clarity} and \textit{test coverage}. 
Task clarity evaluates whether a task description provides sufficient and unambiguous information for solving the problem~\cite{spice, swebenchverified}; incomplete or ambiguous descriptions may lead to incorrect or biased understanding, rendering tasks unsolvable~\cite{mu2024clarifygpt, jia2025automated}. 
Test coverage evaluates whether the test suite is adequate and precise for verifying correctness~\cite{spice, swebenchverified}; insufficient coverage may allow incorrect solutions to pass due to limited discriminative power~\cite{yu2025utboost, wang2025solved}.

Currently, the assessment of task quality in SWE datasets primarily relies on human experts~\cite{swebenchverified, multiswebench, nocode-bench}. 
For example, in SWE-bench Verified~\cite{swebenchverified}, each instance is manually assessed/annotated three times by a total of 93 professional Python engineers. 
While manual annotation can provide relatively high-quality assessments, it is inherently expensive and difficult to scale, especially as large volumes of SWE data are rapidly collected~\cite{SWE-rebench, r2egym} or synthesized~\cite{Swe-mirror, Swe-smith}. 
Specifically, annotating 1,000 instances can cost approximately \$100,000 (\$100 per instance), with an average of 1.05 person-hours required per instance~\cite{spice}. These challenges highlight the need for automated annotation techniques that reduce costs while maintaining annotation quality.

A natural direction is to leverage LLMs for automated SWE dataset annotation. Recent work (\textit{e.g.,} SPICE~\cite{spice}) constructs a rationale checklist and incorporates it into prompts to guide the annotation of SWE instances. However, these judgment rubrics are typically fixed and overly generic, limiting their ability to accommodate diverse and complex scenarios, particularly corner cases. More fundamentally, existing work largely treats annotation as a \textit{static} decision process, whereas prior studies on human annotation and related tasks (\textit{e.g.,} qualitative analyses of annotation behavior~\cite{rottger2022two, schmeisser2022criteria}) suggest that annotation is influenced by contextual factors and \textit{accumulated} experience. This contrast motivates an experience-informed annotation process in which the judgment rubric evolves by learning from previously encountered cases.

However, effectively accumulating such experience is non-trivial. A naive strategy that directly aggregates experience at the instance level can produce increasingly expanded and fragmented rubrics, degenerating into collections of detailed, case-specific rules rather than coherent and generalizable guidelines for decision-making~\cite{hashemi2024llm, siro2026learning}. The key challenge is therefore to balance \textit{generality} and \textit{specificity} when constructing and evolving the judgment rubric. On the one hand, the rubric should capture general principles that transfer across instances; on the other hand, it should retain sufficient specificity to accommodate nuanced and corner cases. 

To this end, we model annotation as an iterative and experience-driven process that progressively refines the judgment rubric through experience abstraction to balance generality and specificity. 
Specifically, we propose \appname, a self-evolving and experience-guided assistant for task quality assessment (\textit{i.e.,} task clarity and test coverage) in SWE datasets.  
\appname is designed for both fully automated annotation and human-assisted annotation, where it provides preliminary assessments to potentially reduce manual effort.
The core of \appname is an \textit{evolving experience bank}, which stores and updates structured knowledge derived from previously processed instances. 
\appname operates in two phases: experience accumulation and experience reuse. The accumulation phase aims to mine cross-instance information from annotated instances into the experience bank, which is iteratively evolved to form a \textit{stable} judgment rubric; the reuse phase then leverages the rubric through experience retrieval when predicting new annotations.

In the accumulation phase, \appname iteratively refines its internal judgment rubric through a three-stage cycle: rollout, reflection, and experience update. 
The rollout stage summarizes each SWE instance along three dimensions (\textit{i.e.,} task type, task explanation, and test explanation), and encodes them using both sparse (\textit{i.e.,} BM25~\cite{BM25}) and dense (LLM-based) representations to enable effective retrieval of relevant past experiences.
The reflection stage leverages the semantic understanding capability of an LLM to compare the predicted annotation with the golden one, prompt the model to \textit{self-reflect} on its reasoning process~\cite{renze2024self}, and identify typical understanding, success, and error patterns, which are then distilled into an experience item. 
Finally, in the experience update stage, the experience bank is dynamically revised (via insertion, modification, and deletion), enabling the model to form a self-evolving cycle that progressively stabilizes the rubric to strike a balance between generality and specificity.

Evaluating such techniques poses an additional challenge, as no publicly available dataset is readily suitable for this purpose~\cite{spice}. To address this, we construct \bench based on the human annotations of SWE-bench~\cite{SWE-bench} and Multi-SWE-bench~\cite{multiswebench}, which consists of a Python subset \benchpy and a multilingual subset \benchml covering seven other programming languages.
Specifically, we retain the instances on which all annotators agree, split them into historical data for experience accumulation and a test set, and apply an additional round of manual cleaning to remove typical annotation errors.
This process results in 1,076/1,907 instances for task clarity assessment and 833/1,739 instances for test coverage assessment in \benchpyml, respectively.

On this benchmark, we evaluate \appname through two assessments using three LLMs, Qwen3.8-27B~\cite{qwen38_27b}, DeepSeek-V4-Flash~\cite{deepseekv4}, and GPT-5.6 Luna~\cite{openai_gpt56_luna}, and compare it with five baselines.
Experimental results show that \appname achieves the best F1 in all 12 combinations of assessment tasks, subsets, and backbone models, with relative improvements in average F1 of 44.2\%/49.9\% and 24.0\%/18.4\% over the best-performing baseline for task clarity and test coverage assessment in \benchpyml, respectively.
Moreover, \appname achieves average FPRs of 3.3\% and 15.1\% for task clarity and test coverage assessment, respectively, indicating that it filters out low-quality instances with limited impact on valid ones.
Ablation studies further demonstrate that removing the experience bank or any of its key components leads to significant performance degradation, highlighting the effectiveness of our design.

Beyond assessment accuracy, we further evaluate whether \appname can improve the quality of SWE data used for coding-agent training. We apply \appname to filter 5k SWE-rebench instances~\cite{SWE-rebench} and fine-tune Qwen3.5-9B-Base~\cite{qwen35} on the resolved trajectories of the retained instances. The resulting model achieves a 66.2\% resolution rate on SWE-bench Verified, outperforming training with data filtered by the best-performing baseline by 14.5\% and with the unfiltered dataset by 21.7\%, while reducing the cost per resolved trajectory to \$0.46.

In summary, our main contributions are as follows:
\begin{itemize}[left=0pt, topsep=0em]
    \item We propose \appname, a novel LLM-based, experience-guided assistant for task quality assessment in SWE datasets. \appname introduces a self-evolving experience bank that accumulates cross-instance knowledge from historical annotations into a stable yet adaptive judgment rubric and reuses relevant experience to guide the assessment of new instances.
    
    \item We construct \bench, a benchmark for task clarity and test coverage assessment spanning eight programming languages, by manually cleaning the human annotations of SWE-bench~\cite{swebenchverified} and Multi-SWE-bench~\cite{multiswebench}.
    
    \item We comprehensively evaluate \appname on \bench, and it achieves the best F1 in all settings with relative improvements in average F1 of 44.2\%/49.9\% and 24.0\%/18.4\% on the two assessments in \benchpyml, respectively, compared with the best-performing baseline.
    
    \item We demonstrate the downstream effectiveness of \appname by applying it to filter 5k SWE-rebench instances~\cite{SWE-rebench} for coding-agent training. Fine-tuning Qwen3.5-9B-Base on the filtered data achieves a 66.2\% resolution rate on SWE-bench Verified, outperforming the best-performing baseline by 14.5\% and the unfiltered data by 21.7\%.

\end{itemize}


\section{Background and Motivation}
\begin{figure*}[h]
    \centering
    \includegraphics[width=\linewidth]{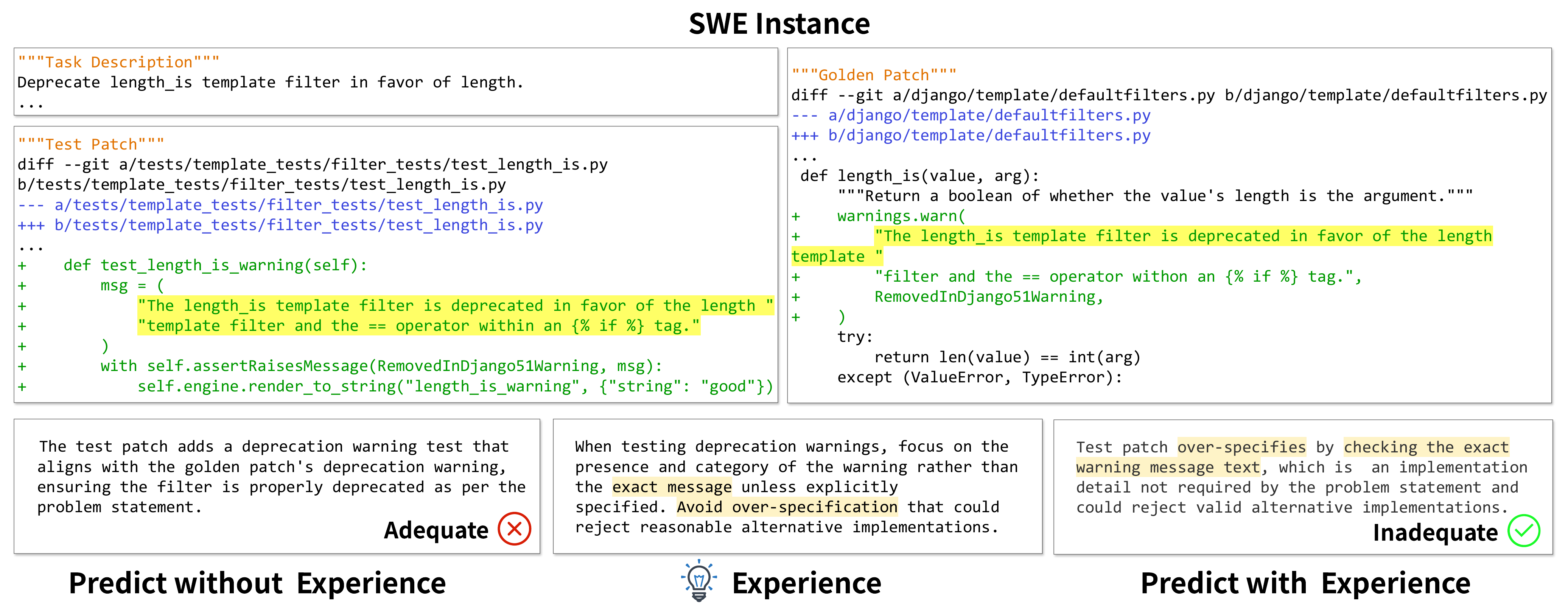}
    \caption{A motivating example of \texttt{django-15869} in SWE-bench.}
    \label{fig: motivation}
\end{figure*}

\subsection{Problem Definition}
\label{sec: problem definition}

In this work, we mainly focus on assessing the task quality in SWE datasets, especially in solvability and verifiability. 
Each SWE instance typically consists of a task description, a corresponding reference solution (golden patch), and a set of test suites used to validate candidate solutions (test patch).
However, not all SWE instances are equally well-defined or verifiable. 
Ambiguous or incomplete task descriptions can make it difficult to determine the intended solution, while inadequate test suites may fail to reliably distinguish correct solutions from incorrect ones~\cite{swebenchverified, wang2025solved, yu2025utboost}. Following prior work~\cite{swebenchverified, nocode-bench, spice}, we assess the SWE instance in two dimensions: \textbf{Task Clarity} and \textbf{Test Coverage}. 
Together, these dimensions aim to determine whether a given SWE instance is sufficiently well-specified and properly verifiable to serve as a meaningful instance for the SWE dataset.

\begin{itemize}[left=0pt, topsep=0em]

\item \textbf{Task Clarity Assessment} evaluates whether the task description provides sufficient and unambiguous information to solve the task, with the golden and test patches serving as references. However, if necessary information can only be inferred from the patches but is absent from the task description, the task is still considered underspecified. External discussions and linked resources are excluded from the assessment~\cite{spice, swebenchverified, nocode-bench}. Given a SWE instance, task clarity assessment assigns a Boolean label, where \texttt{True} indicates that the task is \textit{underspecified} and \texttt{False} indicates that it is \textit{well-specified}, together with a note explaining the reasons and evidence for the judgment.

\item \textbf{Test Coverage Assessment} examines whether the provided test suite is appropriate and sufficient to validate correct solutions to the task. 
Given an instance in a SWE dataset, the test coverage assessment aims to assign a Boolean label mainly according to the test patch, where \texttt{True} indicates that the test suite is \textit{inadequate}, and \texttt{False} indicates that the test suite is \textit{adequate}, with a predicted note to explain the reasons.

\end{itemize}

\subsection{Motivating Example}

Figure~\ref{fig: motivation} presents an instance from the Django project in SWE-bench~\cite{SWE-bench}. 
The task asks to deprecate the \texttt{length\_is} template filter in favor of \texttt{length}. 
To implement this requirement, the golden patch raises a \texttt{RemovedInDjango51Warning} when \texttt{length\_is} is invoked and provides a specific warning message recommending alternative usage. 
The test patch verifies both the warning type and its exact message through \texttt{assertRaisesMessage}.
Human annotators judge this test suite as \textit{inadequate} because the task description requires the filter to be deprecated but does not specify the exact warning message. 
Consequently, a valid solution that raises the correct deprecation warning with different wording may still fail the test. 
Without relevant prior experience, the LLM instead predicts the test suite as \textit{adequate}. 
As illustrated in Figure~\ref{fig: motivation}, experience accumulated from previous annotations can provide a general guideline that tests should avoid imposing implementation details not explicitly required by the task. 
With this experience, the LLM recognizes the exact-message check as over-specification and correctly predicts the test suite as \textit{inadequate}. 
This example motivates us to accumulate experience from historical annotations and reuse relevant experience when assessing new SWE instances.

\section{Approach}
\label{Methodology}

\begin{figure}[t]
    \centering
    \includegraphics[width=1\linewidth]{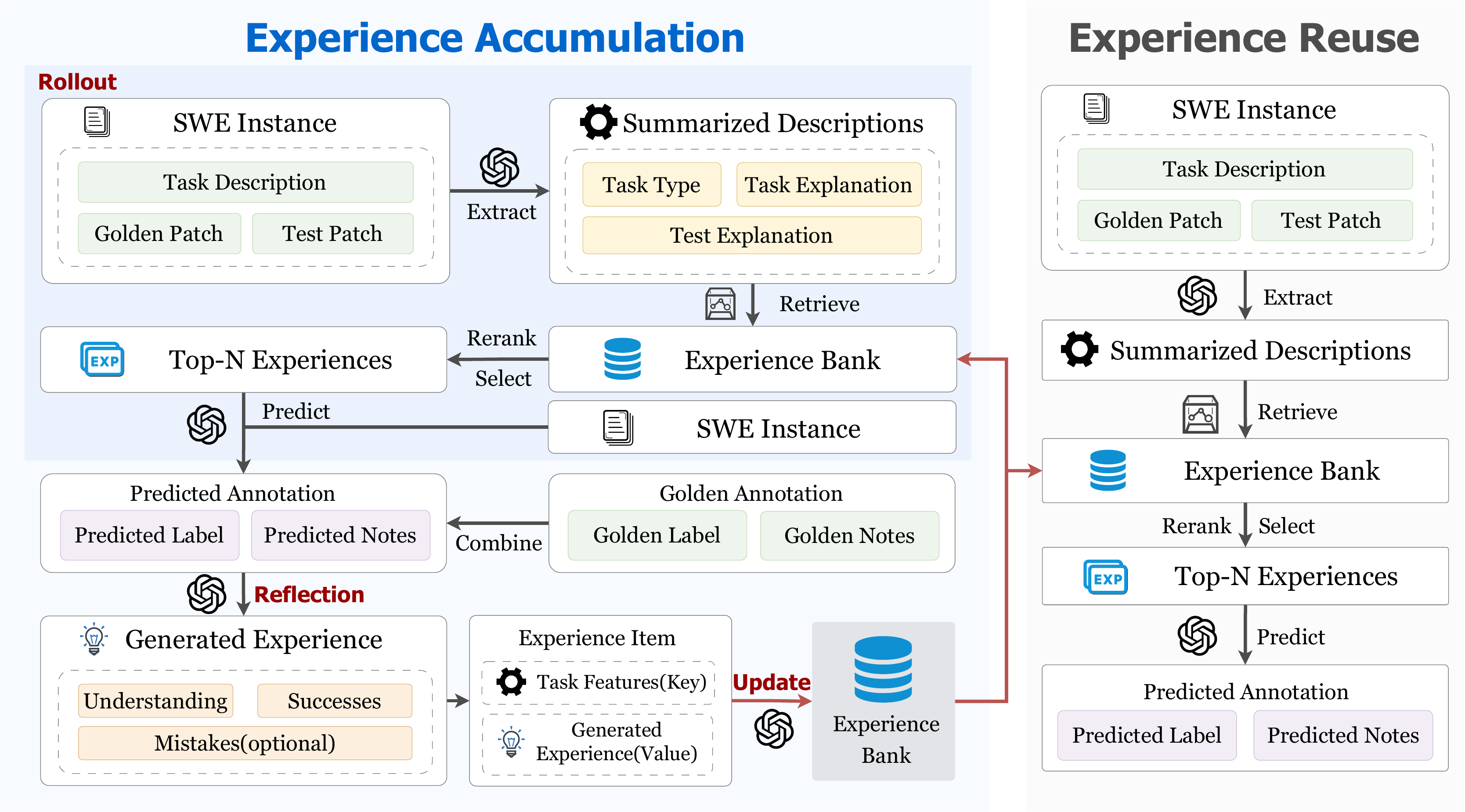}
    \caption{Overview of \appname.}
    \label{fig: overview}
\end{figure}

In this section, we introduce \appname, a self-evolving experience-guided assistant to assess task clarity and test coverage in SWE datasets. 
As illustrated in Figure~\ref{fig: overview}, \appname maintains an experience bank as the central vehicle of self-evolution to accumulate and reuse experience.
Specifically, it consists of two primary phases: experience accumulation and experience reuse.
In the experience accumulation phase (Section~\ref{phase1}), we begin with an empty experience bank and apply an LLM-based (\textit{e.g.,} GPT-5.6 Luna~\cite{openai_gpt56_luna}) retrieval-augmented generation (RAG) method~\cite{racg_survey, RAG} to each labeled instance to predict its annotation.
Subsequently, by comparing the predicted annotation with the golden annotation, we perform self-reflection~\cite{renze2024self, ji2023towards} that guides the LLM to reflect on its reasoning trajectories to generate experience. 
Self-reflection can help the LLM transform concrete RAG-based reasoning trajectories into generalized experience, thereby enabling cross-instance reuse.
To achieve self-evolution, we further manipulate and update the experience bank using three predefined operation interfaces.
A generalized experience bank is finally formed through continuous self-evolution that extracts high-consensus patterns and typical error patterns from a large amount of historical annotation data.
When a new instance arrives, \appname leverages the cross-instance information formed through self-evolution to predict its annotation by reusing the accumulated experiences (Section~\ref{phase2}).

\subsection{Experience Accumulation}\label{phase1}

The experience accumulation phase aims to extract high-consensus patterns, successful annotation strategies, and typical error patterns from annotated instances as experience items that iteratively update the experience bank.
It takes a SWE instance and its golden annotation (\textit{i.e.,} golden label and golden notes) as input and returns an experience item to update the experience bank. 
We design three stages for experience accumulation: rollout, reflection, and experience update. 
Specifically, in the rollout stage (Section~\ref{rollout}), we use an LLM-based RAG pipeline to retrieve experience from the current experience bank and, based on it, predict the annotation of the instance.
In the reflection stage (Section~\ref{reflection}), we compare the current predicted annotation with the golden annotation and apply a self-reflection mechanism to generate experience.
In the experience update stage (Section~\ref{update}), we provide the LLM with an interface for operating on the experience bank, enabling it to insert, modify, or delete the generated experience.

\subsubsection{Rollout}\label{rollout}

This stage aims to generate predicted annotations (\textit{i.e.,} predicted labels and notes) for SWE instances based on the current experience bank, using a RAG-based pipeline. The process can be divided into the following three steps.

\begin{figure}[t]
    \centering
    \includegraphics[width=1\linewidth]{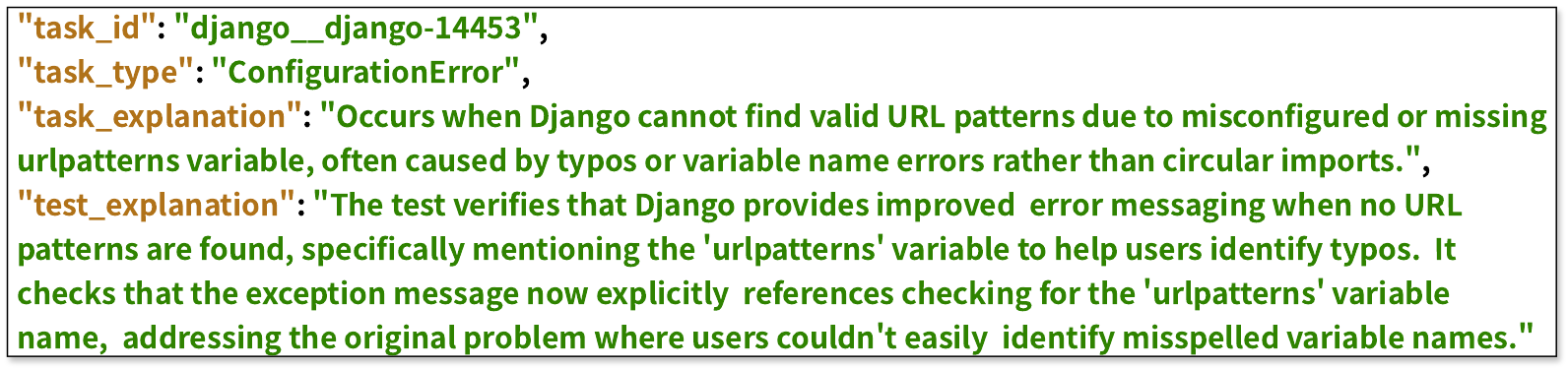}
    \caption{An example of the summarized descriptions along three dimensions for \texttt{django-14453} in SWE-bench.}
    \label{fig: task_features}
\end{figure}

\textbf{Step I: Summarization of the Description in Three Dimensions.}
A prerequisite for forming experience is a structured understanding of SWE instances.
Given that human annotators may interpret task descriptions from multiple perspectives~\cite{lindahl2024disagreement}, we employ an LLM to summarize each SWE instance along multiple dimensions to provide a structured understanding of the instance. 
The dimensions are defined as follows:
\begin{itemize}[left=0pt, topsep=0em]
  \item \textbf{Task Type}: A generalized categorization provided by the LLM using SWE instances (\textit{e.g.,} \texttt{ConfigurationError, Variable-\allowbreak ReferenceError}), which is used to distinguish different task intents. 
  \item \textbf{Task Explanation}: A generalized, concise description of the SWE instance while preserving its core semantics. It describes the underlying causes, trigger conditions, and typical scenarios.
  \item \textbf{Test Explanation}: A summary of the test patch’s logic, validation objectives, and its correspondence with the SWE instance, which is used to distinguish different test behaviors.
\end{itemize}

\noindent These three dimensions correspond to the three components of a SWE instance, and we consider them capable of reflecting the information contained in the SWE instance. 
We first generate descriptions that differentiate among task types, as tasks of the same type may share similar solutions. 
Moreover, we summarize the golden solution and test suites, originally in the code format, of the SWE instances into natural language (NL) descriptions to bridge the semantic gap between code and text. 
The resulting descriptions are concatenated and encoded using an embedding model to form features for indexing experience items (as detailed in \textbf{Step II}).
To help better understand the summarized descriptions, we provide an example shown in Figure~\ref{fig: task_features}.
By constructing and leveraging these descriptions, we achieve a transformation from specific, bimodal SWE instance data to generalized, NL-based data, which not only bridges the semantic gap between text and code but also provides the foundation for generalizing similar experiences and retrieving relevant experience items.

\textbf{Step II: Experience Retrieval.}
This step aims to retrieve, rerank, and select the most relevant experience items from the experience bank.
It allows the LLM to autonomously make predictions based on the existing experience bank, while providing signals for potential refinement of the experience bank.
Following prior work~\cite{ma2025alibaba, lu2022reacc}, we adopt a hybrid strategy that combines sparse retrieval and dense retrieval.
For each SWE instance, we first concatenate the three summarized descriptions generated in Step I to form a unified textual representation. 
We then compute similarity scores between the query instance and each experience item (as detailed in Section~\ref{reflection}) in the current experience bank from both sparse and dense perspectives.
The representation of the three-dimensional descriptions constitutes the \textit{key} (index) of an experience item.
Specifically, we adopt the widely used BM25~\cite{BM25} algorithm for sparse retrieval to capture lexical similarity. 
For dense retrieval, we use an LLM-based embedding model (\textit{e.g.,} Qwen3-Embedding-8B~\cite{qwen3embedding}) to obtain semantically rich representations.
The similarity scores from both retrieval methods are normalized and averaged to produce a final similarity score. Based on this score, experience items are ranked, and the Top-$K$ items are selected.
This hybrid design leverages the complementary strengths of sparse and dense retrieval: BM25 captures lexical similarity~\cite{gao2021complement}, while dense embeddings encode semantic and structural similarity~\cite{karpukhin2020dense}, leading to more accurate and robust experience retrieval.
Furthermore, we introduce an LLM-based reranking step to refine the usability of retrieved experience items. Specifically, the SWE instance and the Top-$K$ retrieved items are provided as context to prompt the LLM, which produces a ranking based on their relevance and usefulness for annotating the current instance. We then select the Top-$N$ (where $N<K$) items from the reranked list as the final retrieval results, which serve as reliable prior knowledge for subsequent prediction generation.

\textbf{Step III: Prediction Generation.}
This step aims to predict the annotation of SWE instances by leveraging the retrieved experience items. 
It takes a SWE instance and the Top-$N$ experience items retrieved in Step II as input and requires the LLM to return the predicted label and predicted notes as predicted annotations.

\begin{figure}[t]
    \centering
    \includegraphics[width=1\linewidth]{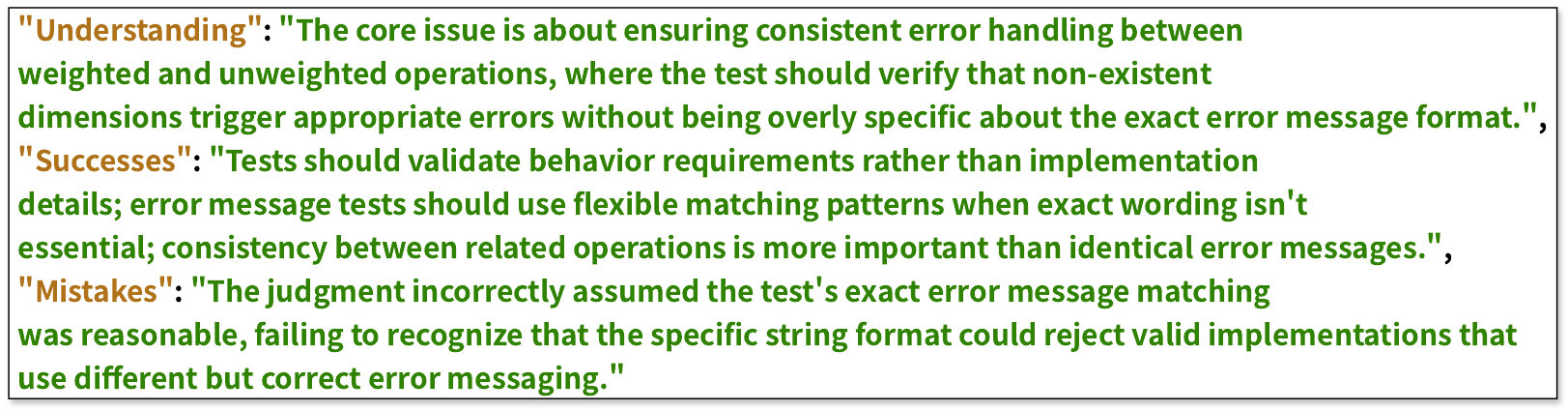}
    \caption{An example of generated experience about \texttt{xarray-5362} in SWE-bench for test coverage assessment.}
    \label{fig: experience}
\end{figure}

\subsubsection{Reflection}\label{reflection}

This stage aims to generate experience items by comparing the predicted annotation with the golden annotation through a self-reflection mechanism. 
Self-reflection is used to reflect on the reasoning process in the rollout stage, which takes the SWE instance, predicted annotation with Top-$N$ retrieved experience items used in the rollout stage, and the golden annotation as input, and returns new experience items for updating the experience bank.
Specifically, we use the LLM to analyze the similarities and differences between the predicted annotation and the golden annotation according to the retrieved experience items in order to generate new experiences, \textit{i.e.,} the transferable and generalizable reasoning patterns.
To capture different ways in which annotation outcomes inform future decisions, we design the \textit{value} of each experience item along three dimensions: Understanding, Successes, and Mistakes. 
These correspond to refined high-consensus understanding, effective annotation strategies, and recurring error patterns, respectively.
The dimensions are defined as follows:
\begin{itemize}[left=0pt, topsep=0em]
    \item \textbf{Understanding}: It is a high-level abstract understanding of the problem.
    It provides a general pattern summarized from the problem and supports cross-instance comparison.  
    \item \textbf{Successes}: It records transferable principles, strategies, and insights distilled from the current instance.
    It provides actionable guidance for how to make successful predictions in the future in similar scenarios. 
    \item \textbf{Mistakes}: It is used to learn from errors when predictions fail. It documents typical patterns such as reasoning flaws, overlooked evidence, and cognitive biases that lead to incorrect outcomes, and is populated only for failed predictions.
\end{itemize}

\noindent Figure~\ref{fig: experience} illustrates an example of experience generated by the LLM for a specific SWE instance (\texttt{xarray-5362}) from SWE-bench~\cite{SWE-bench}. 
For deduplication, each experience item is additionally assigned a unique identifier (\texttt{id}).

\subsubsection{Experience Update}\label{update}

This stage aims to update the experience bank based on newly generated experiences, thereby enabling its self-evolution. 
It takes as input the Top-$N$ retrieved experience items from the rollout stage and the newly generated experience item from the reflection stage, and updates the experience bank through three predefined operations: insertion, modification, and deletion.
The self-evolution process is implemented via an LLM-driven maintenance procedure, which generates an ordered sequence of operations over the experience bank based on the input and executes them accordingly. 
The interfaces corresponding to these operations are described in Table~\ref{table: operations}.

\begin{table}[h]
\centering
\caption{Operations to update the experience bank.}
\label{table: operations}
\setlength{\tabcolsep}{4pt}
\begin{adjustbox}{width=1\columnwidth}
\begin{tabular}{lll}
\toprule
\textbf{Operation} & 
\textbf{Parameters} & 
\textbf{Explanation} \\
\midrule

\textbf{Insert} 
& item
& Add a new experience item and allocate an ID when the current case reveals a novel pattern.
\\

\midrule

\textbf{Modify} 
& ID, experience 
& Refine the value of an existing experience item when prior experience needs clarification or correction.
\\

\midrule

\textbf{Delete} 
& ID 
& Remove an existing experience item when it is contradicted or deemed incorrect.
\\

\bottomrule
\end{tabular}
\end{adjustbox}
\end{table}

Additionally, to ensure that the newly identified experience is explicitly anchored to the current instance and can be independently retrieved in the future, we insert a new experience item generated in the reflection stage, when the operation list contains a \textbf{Modify} but no \textbf{Insert}.
Experience items that reference the same experience share a common \texttt{id}, so that we can deduplicate by this \texttt{id} and keep only one experience item during retrieval.
After all operations are executed, the refreshed experience bank is used for the next round of rollout and reflection, forming a closed loop that continuously improves subsequent predictions.
The detailed prompt template for the experience accumulation phase can be found in our \hyperref[rp]{replication package}.

\subsection{Experience Reuse}\label{phase2}

This phase aims to apply the constructed experience bank to annotate new unlabeled instances, which takes an unlabeled SWE instance as input and returns the predicted label and predicted notes.
It shares the same RAG workflow with the rollout stage in the experience accumulation phase (Section~\ref{rollout}).
Specifically, it first generates three summarized descriptions for a SWE instance and uses their textual representation as a query to retrieve and rerank the Top-$N$ relevant experience items from the accumulated experience bank.
It finally leverages the experience of the selected experience items as contextual information to predict the label and notes of the current instance.

\section{SWE-TQ: A Benchmark for Task Clarity and Test Coverage Assessment}
\label{benchmark}

\subsection{Benchmark Construction}
\label{benchmark construction}

There is currently no publicly available dataset with human annotations for evaluating automated task quality assessment techniques.
SPICE~\cite{spice}, the closest work to ours, only releases SPICE-Bench, which is annotated using its proposed automated technique, while its human-annotated evaluation dataset remains closed-source.
Therefore, we construct \bench, a new benchmark for evaluating automated task clarity and test coverage assessment techniques, following a construction process similar to that of SPICE and aligned with common practices in empirical SE~\cite{bhatia2020study}.
\bench consists of two subsets: \benchpy\footnote{\url{https://cdn.openai.com/introducing-swe-bench-verified/swe-bench-annotation-results.zip}}, built from the human annotations of SWE-bench~\cite{swebenchverified}, and \benchml\footnote{\url{https://github.com/multi-swe-bench/multi-swe-bench/tree/main/docs/manual-verification/annotation-results}}, built from the human annotations of Multi-SWE-bench~\cite{multiswebench}, which covers seven programming languages (\textit{i.e.,} C, C++, Go, Java, JavaScript, Rust, and TypeScript).
Both subsets are constructed through three steps: annotation collection, data splitting, and manual cleaning.

\textbf{Annotation Collection.}
We collect the human annotations of SWE-bench~\cite{swebenchverified} and Multi-SWE-bench~\cite{multiswebench} for \benchpy and \benchml, respectively.
In SWE-bench, each of the 1,699 instances annotated by OpenAI is independently labeled by three annotators, while in Multi-SWE-bench, each of the 2,456 candidate instances is independently labeled by two annotators.
Following their respective annotation guidelines~\cite{swebenchverified, multiswebench}, we convert the original annotations into binary labels.
For both datasets, we apply the same filtering strategy and retain only the instances on which all annotators reach complete agreement on the binary labels.

\textbf{Data Splitting.}
For each subset, we split the instances into historical data for experience accumulation and a test set.
For test coverage assessment, we adopt a temporal split based on commit timestamps, using the most recent 20\% of instances as the test set.
Since test coverage assessment primarily evaluates the \textit{test patch}, which may be affected by the evolving state of the \textit{codebase}, we adopt this more conservative strategy to temporally separate experience accumulation from test instances.
In contrast, task clarity assessment primarily concerns whether the \textit{task description} provides sufficient information to solve the task (Section~\ref{sec: problem definition}) and is less subject to such temporal dependencies.
Given the class imbalance in task clarity assessment, a 20\% temporal split would leave very few underspecified instances in the test set (10 for \benchpy and 24 for \benchml).
Therefore, we randomly sample 30\% of the instances as the test set to include a sufficient number of underspecified instances for evaluation.

\textbf{Manual Cleaning.}
To further ensure dataset reliability, two authors manually review all instances in \benchpy and \benchml to identify \textit{two} recurring annotation patterns that are inconsistent with OpenAI's annotation criteria~\cite{swebenchverified}.
For task clarity assessment, we remove instances labeled as well-specified when key information (\textit{e.g.,} error messages) is provided only through \textit{external links} excluded by the annotation criteria.
Allowing LLMs to access such external resources may introduce data leakage or benchmark contamination~\cite{choi2025contaminated, deng2024investigating}, unpredictable environmental noise~\cite{mohammadi2025evaluation, potamitis2025reasonbench}, and reproducibility issues~\cite{gargreal}.
For test coverage assessment, we remove instances labeled as adequate when the test patch assumes an exact error message that is not specified in the task description.
We remove these instances rather than relabeling them because each annotation contains both a label and notes justifying the label: changing the label alone would make it inconsistent with the original notes, while rewriting the notes would introduce the authors' own judgments into the human annotations, potentially biasing experience accumulation and evaluation.
More details are provided in Section~\ref{Dataset Re-annotation}.
Instances for which the two authors fail to reach consensus are also removed.
The two authors achieve a Cohen's kappa~\cite{cohen1960coefficient} of 0.928 on their cleaning decisions, indicating strong agreement~\cite{landis1977measurement}.
Overall, manual cleaning removes 34 and 148 instances from \benchpy and 136 and 384 instances from \benchml for task clarity and test coverage assessment, respectively.

\subsection{Benchmark Characteristics}
\label{benchmark characteristics}

\begin{table}[t]
\centering
\caption{Statistics of \bench, where Clarity and Coverage denote the two assessment tasks, Accum. denotes the historical data for experience accumulation, and Desc. denotes the task description. Instance characteristics are averaged over each group.}
\label{table: benchmark statistics}
\begin{adjustbox}{width=\textwidth}
\begin{tabular}{llcccccccccccccc}
\toprule
\multirow{3}{*}{\textbf{Task}} & \multirow{3}{*}{\textbf{Label}}
& \multicolumn{7}{c}{\textbf{\textit{\benchpy}}}
& \multicolumn{7}{c}{\textbf{\textit{\benchml}}} \\
\cmidrule(lr){3-9} \cmidrule(lr){10-16}
& & \multicolumn{2}{c}{\textbf{\#Inst.}} & \textbf{Desc.}
& \multicolumn{2}{c}{\textbf{Golden Patch}} & \multicolumn{2}{c}{\textbf{Test Patch}}
& \multicolumn{2}{c}{\textbf{\#Inst.}} & \textbf{Desc.}
& \multicolumn{2}{c}{\textbf{Golden Patch}} & \multicolumn{2}{c}{\textbf{Test Patch}} \\
\cmidrule(lr){3-4} \cmidrule(lr){5-5} \cmidrule(lr){6-7} \cmidrule(lr){8-9}
\cmidrule(lr){10-11} \cmidrule(lr){12-12} \cmidrule(lr){13-14} \cmidrule(lr){15-16}
& & Accum. & Test & \#Words & \#Files & \#Lines & \#Files & \#Lines
& Accum. & Test & \#Words & \#Files & \#Lines & \#Files & \#Lines \\
\midrule
\multirow{2}{*}{Clarity}
& Underspecified & 43 & 18 & 155.7 & 2.9 & 74.4 & 1.8 & 67.0 & 61 & 26 & 115.5 & 3.7 & 98.0 & 2.0 & 84.2 \\
& Well-specified & 733 & 282 & 201.7 & 1.5 & 26.1 & 1.5 & 31.6 & 1,273 & 547 & 184.4 & 4.6 & 139.5 & 2.7 & 90.6 \\
\midrule
\multirow{2}{*}{Coverage}
& Inadequate & 265 & 56 & 196.2 & 2.2 & 78.2 & 1.8 & 52.0 & 396 & 62 & 174.9 & 4.4 & 108.9 & 2.3 & 68.9 \\
& Adequate & 403 & 109 & 187.1 & 1.2 & 14.2 & 1.3 & 23.2 & 995 & 286 & 176.8 & 4.7 & 134.0 & 2.4 & 82.4 \\
\bottomrule
\end{tabular}
\end{adjustbox}
\end{table}

Table~\ref{table: benchmark statistics} summarizes the statistics of \bench, which spans eight programming languages.
After filtering, \benchpy and \benchml comprise 1,076 and 1,907 instances for task clarity assessment and 833 and 1,739 instances for test coverage assessment, respectively.
Both subsets are imbalanced, especially for task clarity assessment.
In \benchpy and \benchml, only 6.0\% and 4.5\% of the test instances are underspecified, and 33.9\% and 17.8\% are inadequate, respectively.
Such imbalance makes it challenging to precisely identify the rare positive instances without discarding many valid ones.

We further examine whether the reported surface characteristics can distinguish positive and negative instances. Using any single characteristic (\textit{e.g.,} \#Words) yields an AUC of 0.50--0.77, suggesting that effective assessment cannot rely solely on individual surface characteristics and requires considering the semantic information in the task description and patches. 
We assess the adequacy of our test-set sizes using the minimum sample size required under random sampling with a 95\% confidence level and a 10\% margin of error. The resulting minimum sample sizes for task clarity and test coverage assessment are 88/91 and 86/91 for \benchpyml, respectively. 
Our test sets contain 300/573 instances for task clarity assessment and 165/348 instances for test coverage assessment in \benchpyml, respectively, all exceeding the corresponding minimum sample sizes.


\section{Experiment Setup}

\subsection{Research Questions}

\begin{itemize}[left=0pt, topsep=0em]
    \item \textbf{RQ1 (Effectiveness):} How effective is \appname in assessing task clarity and test coverage compared with existing automated techniques?
    \item \textbf{RQ2 (Ablation):} How do the key components of \appname contribute to its overall effectiveness?
    \item \textbf{RQ3 (Downstream Effectiveness):} Can the dataset filtering performed by \appname benefit the training of coding agents?
    \item \textbf{RQ4 (Case Study):} How does the experience of \appname help it perform better?
\end{itemize}

\subsection{Baselines}

We mainly compare \appname with the open-source state-of-the-art baseline method, SPICE~\cite{spice}. 
Additionally, we include direct prompting (\textit{i.e.,} Vanilla) and a simple RAG approach as baselines to highlight the effectiveness of our self-evolving mechanism.
We further include two state-of-the-art general-purpose coding agents, Codex~\cite{OpenAI_codex_2026} and Claude Code~\cite{Anthropic_claudecode_2026}, to examine whether autonomous exploration of the repository is sufficient for assessing task quality.
The implementation details of baselines are described in Section~\ref{details}.
\begin{itemize}[left=0pt, topsep=0em]
    \item \textbf{Vanilla}:  
    The LLM directly predicts the annotation using only the SWE instance without any additional context, which serves as a lower-bound reference for zero-shot performance.
    \item \textbf{SimpleRAG}: 
    The LLM predicts the annotation with retrieval-augmented generation, where the historical annotated instances are stored as external memory and the three most similar ones are retrieved as additional context for each SWE instance.
    \item \textbf{SPICE}: 
    For task clarity assessment, SPICE utilizes \textit{developer-provided} reasoning patterns as guidance for prediction. For test coverage assessment, it leverages Aider~\cite{AiderAI_aider_2026}, an open-source command-line AI programming agent, to incorporate repository-aware code as additional context.
    \item \textbf{Codex \& Claude Code}:
    General-purpose coding agents developed by OpenAI and Anthropic, respectively, which autonomously explore the repository and read relevant code to predict the annotation of each SWE instance.

\end{itemize}

\subsection{Evaluation Metrics}
Since the goal of both tasks is to identify unsolvable or unverifiable instances, following prior detection work~\cite{zhu2025mcp,chen2025jailbreak}, we regard such instances as positive and adopt the following metrics.

\begin{itemize}[left=0pt, topsep=0em]
\item \underline{\textbf{F1-score:}} The harmonic mean of precision and recall, where precision measures the proportion of predicted positive instances that are truly positive, and recall measures the proportion of positive instances that are correctly identified. We adopt the F1-score as the primary metric because it balances precision and recall when identifying positive instances.
\item \textbf{False Positive Rate (FPR):} The proportion of negative instances incorrectly predicted as positive. Since instances predicted as positive are filtered out, FPR measures the proportion of valid (\textit{i.e.,} solvable or verifiable) instances that are mistakenly discarded and thus reflects the data loss caused by filtering.
\end{itemize}

In addition, we report the average number of tokens (\textbf{\#Token}) and monetary cost (\textbf{\$Cost}) consumed per instance to measure the efficiency of each technique.
The corresponding precision and recall values for all techniques are further reported in \hyperref[rp]{\textbf{online Appendix~A}}.

\subsection{Implementation Details}\label{details}

We evaluate all techniques on the test data of both \benchpy and \benchml described in Section~\ref{benchmark}, and \appname accumulates experience on the merged historical data of both subsets for each task, which also serves as the retrieval pool of SimpleRAG.
We further analyze how the size of the historical data affects the effectiveness in \hyperref[rp]{\textbf{online Appendix~B}}.
We use Qwen3.8-27B~\cite{qwen38_27b}, DeepSeek-V4-Flash~\cite{deepseekv4}, and GPT-5.6 Luna~\cite{openai_gpt56_luna} as the backbone models for all methods, as they represent models from different sources (\textit{i.e.,} open-source or closed-source) and families (\textit{i.e.,} dense or MoE).
We use Qwen3-Embedding-8B~\cite{qwen3embedding} to generate dense representations for retrieval, as it has demonstrated state-of-the-art performance across many information retrieval tasks~\cite{mteb}.
This model is used in both \appname and SimpleRAG for computing embeddings to ensure a fair comparison.
All backbone models are accessed through their official APIs.

For \appname, during experience retrieval (Section~\ref{rollout}), we first retrieve the Top-10 candidate experience items, then rerank them to select the Top-3 as the final result, following prior experience retrieval work~\cite{li2026seam,kwon2026memento}.
Since the experience bank is initially empty during the accumulation phase (Section~\ref{phase1}), \appname makes predictions without experience when no items are retrieved.
The experience generated from this prediction is then inserted into the experience bank as a cold start.
For SimpleRAG, we concatenate the task description, golden patch, and test patch of each instance as its textual representation, retrieve the three most similar historical instances by cosine similarity using the same embedding model as \appname, and use them together with their annotations as few-shot examples for prediction.
To facilitate reproducibility, we uniformly set $\texttt{temperature} = 0$ for annotation prediction across all approaches.

\subsection{Downstream Effectiveness Analysis}\label{downstream}

We evaluate the downstream impact of dataset filtering by comparing coding models trained on data retained by different task quality assessment techniques for software issue resolution.
Specifically, we first randomly sample 5k instances from SWE-rebench~\cite{SWE-rebench}, a large-scale SWE dataset constructed through automated data collection, and sample the corresponding trajectories using DeepSeek-V4-Flash~\cite{deepseekv4}.
To ensure a fair comparison, we apply \appname and all baselines with GPT-5.6 Luna as the backbone model to filter out each instance predicted as either underspecified or inadequate, together with its trajectories.
For \appname, we directly reuse the experience banks accumulated with GPT-5.6 Luna in RQ1.
We then perform supervised fine-tuning (SFT) on Qwen3.5-9B-Base~\cite{qwen35} using the resolved trajectories of the remaining instances and evaluate the resulting models on SWE-bench Verified~\cite{swebenchverified}.
The detailed training recipe is provided in \hyperref[rp]{\textbf{online Appendix~C}}.
We also include the base model without fine-tuning (Base Model) and the model trained on all 5k instances without filtering (Full Data) as references.
For each technique, we report the number of filtered instances (\#Filtered), the number of remaining resolved trajectories (\#Trajs), the annotation cost (\$Annotation), the sampling cost (\$Sampling), and their total cost per trajectory (\$Cost/\#Trajs), and adopt the resolution rate on SWE-bench Verified (Resolved\%) as the primary metric.
Note that \$Sampling is calculated only for the instances retained by each technique, reflecting the expected sampling cost when filtering is performed before trajectory sampling.


\section{Experimental Results}

\subsection{RQ1: Effectiveness}

\begin{table*}[t!]
\centering
\caption{Overall performance on the task clarity and test coverage assessments, where Python and Multilingual denote \benchpy and \benchml, respectively, and Avg. Impr. (F1) denotes the relative improvement of the average F1 of \appname across the three backbone models over that of each baseline.}
\label{full}
\setlength{\tabcolsep}{4pt}
\begin{adjustbox}{width=\textwidth}
\begin{tabular}{cclcccc|cccc|cccc|c}
\toprule
& & \multirow{2}{*}{\textbf{Method}}
& \multicolumn{4}{c}{\textbf{Qwen3.8-27B}} & \multicolumn{4}{c}{\textbf{DeepSeek-V4-Flash}} & \multicolumn{4}{c}{\textbf{GPT-5.6 Luna}} & \multirow{2}{*}{\makecell{\textbf{Avg. Impr.}\\ \textbf{(F1)}}} \\
\cmidrule(lr){4-7} \cmidrule(lr){8-11} \cmidrule(lr){12-15}
& & & \textbf{F1}$\uparrow$ & FPR$\downarrow$ & \#Token & \$Cost & \textbf{F1}$\uparrow$ & FPR$\downarrow$ & \#Token & \$Cost & \textbf{F1}$\uparrow$ & FPR$\downarrow$ & \#Token & \$Cost & \\
\midrule
 &  & Vanilla & 50.0 & \textbf{0.0} & 1.1k & 0.0010 & 24.7 & 16.3 & 0.9k & 0.0002 & 45.5 & 5.7 & 1.1k & 0.0004 & \upp{47.8} \\
 &  & SimpleRAG & 43.5 & \textbf{0.0} & 4.4k & 0.0027 & 19.6 & 36.2 & 4.1k & 0.0009 & 48.7 & 3.6 & 4.2k & 0.0009 & \upp{58.9} \\
 &  & SPICE & 52.8 & 7.4 & 2.4k & 0.0030 & 39.6 & 19.5 & 1.5k & 0.0005 & 30.8 & 28.7 & 1.5k & 0.0006 & \upp{44.2} \\
 &  & Codex & 20.0 & \textbf{0.0} & 168.8k & 0.0740 & 30.3 & 3.6 & 47.5k & 0.0016 & 52.6 & 3.6 & 135.1k & 0.0102 & \upp{72.7} \\
 &  & Claude Code & 20.0 & \textbf{0.0} & 103.4k & 0.0230 & 37.2 & 6.0 & 92.0k & 0.0208 & 47.6 & 5.0 & 80.4k & 0.0063 & \upp{69.6} \\
\rowcolor{paleGreen}
\cellcolor{white} & \cellcolor{white}\multirow{-6}{*}{\rotatebox[origin=c]{90}{Python}} & \appnamebold & \textbf{64.7} & 1.8 & 7.3k & 0.0068 & \textbf{51.9} & \textbf{0.7} & 9.7k & 0.0028 & \textbf{61.1} & \textbf{2.5} & 6.0k & 0.0021 & -- \\
\cmidrule(lr){2-16}
 &  & Vanilla & 26.7 & 4.8 & 1.1k & 0.0011 & 15.0 & 26.9 & 0.8k & 0.0002 & 17.3 & 20.3 & 1.0k & 0.0003 & \upp{52.2} \\
 &  & SimpleRAG & 26.9 & 3.5 & 3.7k & 0.0025 & 12.2 & 51.7 & 3.2k & 0.0008 & 20.8 & 11.0 & 3.5k & 0.0010 & \upp{49.9} \\
 &  & SPICE & 13.9 & 22.1 & 2.1k & 0.0027 & 15.4 & 32.5 & 1.7k & 0.0006 & 13.2 & 44.4 & 1.4k & 0.0006 & \upp{111.3} \\
 &  & Codex & 17.1 & 1.1 & 105.0k & 0.0184 & 16.6 & 19.6 & 25.6k & 0.0013 & 14.5 & 16.3 & 158.0k & 0.0120 & \upp{86.3} \\
 &  & Claude Code & 24.2 & \textbf{0.6} & 144.7k & 0.0681 & 16.2 & 11.9 & 34.7k & 0.0079 & 13.3 & 21.0 & 98.1k & 0.0077 & \upp{67.2} \\
\rowcolor{paleGreen}
\cellcolor{white}\multirow{-12}{*}{\rotatebox[origin=c]{90}{Task Clarity}} & \cellcolor{white}\multirow{-6}{*}{\rotatebox[origin=c]{90}{Multilingual}} & \appnamebold & \textbf{37.5} & 2.4 & 6.8k & 0.0080 & \textbf{25.4} & \textbf{6.6} & 6.8k & 0.0023 & \textbf{26.9} & \textbf{5.9} & 7.9k & 0.0021 & -- \\
\midrule
 &  & Vanilla & 10.0 & 0.9 & 4.9k & 0.0067 & 13.3 & \textbf{0.0} & 2.8k & 0.0007 & 12.7 & \textbf{2.8} & 3.5k & 0.0016 & \upp{535.8} \\
 &  & SimpleRAG & 66.7 & 1.8 & 13.1k & 0.0089 & 31.9 & 1.8 & 10.6k & 0.0024 & 20.9 & 3.7 & 11.5k & 0.0035 & \upp{91.5} \\
 &  & SPICE & 71.3 & 24.8 & 42.2k & 0.0333 & 37.5 & 8.3 & 44.2k & 0.0102 & 75.8 & 23.9 & 42.0k & 0.0085 & \upp{24.0} \\
 &  & Codex & 10.0 & 0.9 & 265.5k & 0.1220 & 16.9 & 8.3 & 48.0k & 0.0027 & 43.6 & 4.6 & 153.3k & 0.0118 & \upp{224.7} \\
 &  & Claude Code & 0.0 & \textbf{0.0} & 158.6k & 0.0408 & 9.2 & 5.5 & 35.3k & 0.0084 & 30.6 & 4.6 & 99.8k & 0.0089 & \upp{475.1} \\
\rowcolor{paleGreen}
\cellcolor{white} & \cellcolor{white}\multirow{-6}{*}{\rotatebox[origin=c]{90}{Python}} & \appnamebold & \textbf{76.5} & 6.4 & 15.7k & 0.0191 & \textbf{73.0} & 15.6 & 21.0k & 0.0065 & \textbf{79.4} & 21.1 & 13.6k & 0.0072 & -- \\
\cmidrule(lr){2-16}
 &  & Vanilla & 24.4 & 3.5 & 5.4k & 0.0066 & 33.7 & 7.7 & 3.4k & 0.0008 & 20.3 & \textbf{3.2} & 4.4k & 0.0021 & \upp{42.6} \\
 &  & SimpleRAG & 26.2 & 15.4 & 9.8k & 0.0117 & 13.3 & \textbf{2.8} & 12.0k & 0.0027 & 36.0 & 7.0 & 11.9k & 0.0028 & \upp{48.1} \\
 &  & SPICE & 31.7 & 37.8 & 28.9k & 0.0299 & 27.0 & 21.0 & 22.2k & 0.0054 & 35.7 & 47.6 & 23.9k & 0.0055 & \upp{18.4} \\
 &  & Codex & 14.1 & \textbf{1.4} & 225.9k & 0.0461 & 19.1 & 11.5 & 29.7k & 0.0023 & 38.9 & 8.7 & 220.3k & 0.0157 & \upp{55.1} \\
 &  & Claude Code & 16.7 & \textbf{1.4} & 260.1k & 0.1318 & 8.9 & 8.4 & 47.5k & 0.0160 & 35.2 & 9.4 & 128.1k & 0.0111 & \upp{83.9} \\
\rowcolor{paleGreen}
\cellcolor{white}\multirow{-12}{*}{\rotatebox[origin=c]{90}{Test Coverage}} & \cellcolor{white}\multirow{-6}{*}{\rotatebox[origin=c]{90}{Multilingual}} & \appnamebold & \textbf{36.9} & 7.7 & 15.7k & 0.0185 & \textbf{34.9} & 9.8 & 13.5k & 0.0031 & \textbf{40.0} & 30.1 & 16.6k & 0.0093 & -- \\
\bottomrule
\end{tabular}
\end{adjustbox}
\end{table*}

The experimental results of all techniques on task clarity and test coverage assessment using three LLM backbones (\textit{i.e.,} Qwen3.8-27B, DeepSeek-V4-Flash, and GPT-5.6 Luna) are reported in Table~\ref{full}.
Compared with the five baselines, \appname achieves the best F1 across all 12 combinations of assessment tasks, subsets, and backbone models.
Moreover, averaged across the three backbone models, \appname improves F1 over the best-performing baseline by 44.2\%/49.9\% for task clarity assessment and 24.0\%/18.4\% for test coverage assessment on \benchpyml, respectively.
These consistent improvements across the three backbone models demonstrate the effectiveness of leveraging cross-instance information through the evolving experience bank for identifying unsolvable and unverifiable instances in SWE datasets.

Although several baselines achieve lower FPRs than \appname in some settings, they often do so while achieving substantially lower F1.
For example, Claude Code with Qwen3.8-27B achieves an FPR of 0.0\% on test coverage assessment in \benchpy, but its F1 is also 0.0\%.
SPICE, in contrast, exhibits relatively high FPRs in several settings, reaching 47.6\%, indicating that it can mistakenly filter out a substantial proportion of valid instances.
This may stem from its reliance on fixed rubrics, which are less adaptive to the nuanced differences across instances.
In comparison, \appname achieves a higher F1 than SPICE in all 12 combinations while maintaining a lower FPR in 11 of them.
Moreover, although Codex and Claude Code can autonomously explore repositories, their F1 reaches at most 52.6\%, suggesting that repository exploration alone does not ensure accurate task quality assessment.
Together, these results show that the evolving experience bank enables \appname to more effectively distinguish unsolvable or unverifiable instances from valid ones than fixed-rubric and general-purpose coding-agent baselines.
In terms of cost, \appname consumes 6.0k--21.0k tokens per instance, substantially fewer than the 25.6k--265.5k tokens consumed by Codex and Claude Code, suggesting that reusing accumulated experience can reduce the need for extensive repository exploration.

Despite its higher F1, \appname still incurs FPRs of 0.7\%--6.6\% and 6.4\%--30.1\% for task clarity and test coverage assessment, respectively, meaning that some valid instances are inevitably removed together with the identified positive instances.
However, such false positives affect the quantity of retained data rather than introducing unsolvable or unverifiable instances into it.
This trade-off motivates our downstream evaluation in Section~\ref{rq3}, where we directly examine whether filtering with \appname benefits subsequent coding-agent training despite the loss of some valid instances.

\begin{table*}[t!]
\centering
\caption{Ablation study results with GPT-5.6 Luna, where Python and Multilingual denote \benchpy and \benchml, respectively, and F1 and FPR are reported in percentage.}
\label{table: ablation}
\setlength{\tabcolsep}{4pt}
\begin{adjustbox}{width=\textwidth}
\begin{tabular}{clcccc|cccc}
\toprule
& \multirow{2}{*}{\textbf{Method}}
& \multicolumn{4}{c}{\textbf{Python}} & \multicolumn{4}{c}{\textbf{Multilingual}} \\
\cmidrule(lr){3-6} \cmidrule(lr){7-10}
& & \textbf{F1}$\uparrow$ & FPR$\downarrow$ & \#Token & \$Cost & \textbf{F1}$\uparrow$ & FPR$\downarrow$ & \#Token & \$Cost \\
\midrule
\rowcolor{paleGreen}
\cellcolor{white} & \appnamebold & \textbf{61.1} & \textbf{2.5} & 6.0k & 0.0021 & \textbf{26.9} & \textbf{5.9} & 7.9k & 0.0021 \\
 & \hspace{1em}w/o Experience Bank & 45.5\downp{25.5} & 5.7 & 1.1k & 0.0004 & 17.3\downp{35.7} & 20.3 & 1.0k & 0.0003 \\
 & \hspace{1em}w/o Reranking & 48.8\downp{20.1} & 4.6 & 2.5k & 0.0010 & 17.7\downp{34.2} & 14.1 & 2.4k & 0.0010 \\
 & \hspace{1em}w/o Field (Successes) & 53.7\downp{12.1} & 4.3 & 4.3k & 0.0013 & 22.2\downp{17.5} & 11.3 & 4.5k & 0.0017 \\
 & \hspace{1em}w/o Field (Mistakes) & 46.5\downp{23.9} & 5.3 & 5.3k & 0.0019 & 20.2\downp{24.9} & 11.5 & 5.1k & 0.0019 \\
\multirow{-6}{*}{\rotatebox[origin=c]{90}{Task Clarity}} & \hspace{1em}w/o Field (Successes, Mistakes) & 44.9\downp{26.5} & 7.1 & 4.6k & 0.0017 & 18.9\downp{29.7} & 12.8 & 4.4k & 0.0017 \\
\midrule
\rowcolor{paleGreen}
\cellcolor{white} & \appnamebold & \textbf{79.4} & 21.1 & 13.6k & 0.0072 & \textbf{40.0} & 30.1 & 16.6k & 0.0093 \\
 & \hspace{1em}w/o Experience Bank & 12.7\downp{84.0} & \textbf{2.8} & 3.5k & 0.0016 & 20.3\downp{49.3} & \textbf{3.2} & 4.4k & 0.0021 \\
 & \hspace{1em}w/o Reranking & 74.1\downp{6.7} & 26.6 & 8.9k & 0.0054 & 38.3\downp{4.3} & 37.4 & 11.0k & 0.0069 \\
 & \hspace{1em}w/o Field (Successes) & 76.0\downp{4.3} & 22.0 & 8.7k & 0.0023 & 38.3\downp{4.3} & 31.5 & 14.7k & 0.0082 \\
 & \hspace{1em}w/o Field (Mistakes) & 76.9\downp{3.1} & 22.0 & 9.2k & 0.0024 & 38.0\downp{5.0} & 33.6 & 14.4k & 0.0074 \\
\multirow{-6}{*}{\rotatebox[origin=c]{90}{Test Coverage}} & \hspace{1em}w/o Field (Successes, Mistakes) & 72.0\downp{9.3} & 22.0 & 8.5k & 0.0023 & 37.1\downp{7.3} & 33.6 & 14.3k & 0.0080 \\
\bottomrule
\end{tabular}
\end{adjustbox}
\end{table*}

\answerRQ{
Compared with the baselines, \appname achieves the best F1 in all settings and improves the average F1 over the best-performing baseline by 44.2\%/49.9\% on task clarity assessment and 24.0\%/18.4\% on test coverage assessment in \benchpyml, respectively.}

\subsection{RQ2: Ablation}

We conduct ablation studies on both task clarity and test coverage assessments by removing key components or experience fields from the \appname design. 
Specifically, we evaluate the following variants: w/o Experience Bank, w/o Reranking, w/o Field (Successes, Mistakes), w/o Field (Successes), and w/o Field (Mistakes). 
All experiments are performed using GPT-5.6 Luna to assess the contribution of each component.
As shown in Table~\ref{table: ablation}, \appname consistently outperforms all five variants in terms of F1, highlighting the effectiveness of its component design.
\begin{itemize}[left=0pt, topsep=0em]
    \item \textbf{w/o Experience Bank}: 
    Removing the experience bank reduces F1 by 25.5\%--35.7\% on task clarity assessment and 49.3\%--84.0\% on test coverage assessment.
    Although this variant yields a lower FPR on test coverage assessment, it identifies far fewer inadequate instances.
    This suggests that without prior experience, the LLM lacks cross-instance guidance and is more prone to incorrect reasoning.
    
    \item \textbf{w/o Reranking}: 
    Removing reranking lowers F1 by 20.1\%--34.2\% and 4.3\%--6.7\% on the two tasks, respectively.
    This is likely because less precise retrieval introduces irrelevant context, increasing noise and distracting the LLM from the core issue.
    
    \item \textbf{w/o Field (Successes)}: 
    Removing the \textbf{Successes} field leads to F1 drops of 12.1\%--17.5\% and 4.3\% on the two tasks, respectively.
    This indicates that \textbf{Successes} provides useful reasoning patterns that guide correct predictions.
    
    \item \textbf{w/o Field (Mistakes)}: 
    Removing the \textbf{Mistakes} field decreases F1 by 23.9\%--24.9\% and 3.1\%--5.0\% on the two tasks, respectively.
    This suggests that \textbf{Mistakes} helps the LLM avoid recurring reasoning errors by highlighting common pitfalls.
    
    \item \textbf{w/o Field (Successes, Mistakes)}: 
    Removing both \textbf{Successes} and \textbf{Mistakes} reduces F1 by 26.5\%--29.7\% and 7.3\%--9.3\% on the two tasks, respectively.
    This shows that the two fields play complementary roles, jointly providing positive guidance and error correction.
\end{itemize}

These results demonstrate that each component of \appname plays an indispensable role in enabling experience extraction with self-reflection and the self-evolution of the experience bank, and they constitute essential elements of \appname’s effectiveness.
We further present a representative case in \hyperref[rp]{\textbf{online Appendix~D}} to illustrate how each component affects the prediction.

\answerRQ{
All components of \appname are effective, and the experience bank is the most critical one, whose removal reduces F1 by 48.6\% on average.
}

\subsection{RQ3: Downstream Effectiveness}\label{rq3}

\begin{table}[t]
\centering
\caption{Downstream effectiveness on the 5k SWE-rebench instances.}
\label{table: application}
\begin{adjustbox}{width=0.9\textwidth}
\begin{tabular}{lccccc|c}
\toprule
\textbf{Method} & \textbf{\#Filtered} & \textbf{\#Trajs} & \textbf{\$Annotation} & \textbf{\$Sampling} & \textbf{\$Cost/\#Trajs} & \textbf{Resolved\%}$\uparrow$ \\
\midrule
Base Model & -- & -- & -- & -- & -- & 2.0 \\
Full Data & 0 & 2,309 & 0.00 & 1,680.94 & 0.73 & 54.4\upp{21.7} \\
\midrule
Vanilla & 2,140 & 1,424 & 9.87 & 876.93 & 0.62 & 51.0\upp{29.8} \\
SimpleRAG & 2,248 & 1,390 & 24.65 & 860.55 & 0.64 & 57.8\upp{14.5} \\
SPICE & 4,124 & 582 & 38.42 & 225.61 & \textbf{0.45} & 57.6\upp{14.9} \\
Codex & 2,471 & 1,361 & 85.04 & 744.58 & 0.61 & 57.4\upp{15.3} \\
Claude Code & 2,465 & 1,330 & 57.46 & 738.05 & 0.60 & 55.6\upp{19.1} \\
\rowcolor{paleGreen}
\appnamebold & 3,276 & 1,087 & 35.21 & 460.54 & 0.46 & \textbf{66.2} \\
\bottomrule
\end{tabular}
\end{adjustbox}
\end{table}

As shown in Table~\ref{table: application}, the model trained on the data filtered by \appname achieves the highest resolution rate of 66.2\%, outperforming the best-performing baseline by 14.5\%.
Notably, \appname retains 1,087 resolved trajectories, fewer than the 1,330--1,424 retained by all baselines except SPICE.
Compared with Full Data, \appname improves the resolution rate by 21.7\% while using fewer than half of its 2,309 trajectories.
These results show that the benefit of \appname does not arise from using more training data; instead, its filtering enables more effective training with substantially fewer trajectories.

Moreover, filtering with \appname reduces the cost per resolved trajectory to \$0.46, comparable to the \$0.45 of SPICE and lower than the \$0.60--\$0.64 of the other baselines, as filtering before sampling avoids spending sampling cost on instances that would subsequently be filtered out.
SPICE achieves a slightly lower cost of \$0.45 but filters out 4,124 of the 5k instances and retains only 582 resolved trajectories.
Such aggressive filtering is consistent with its substantially higher FPR observed in RQ1.
In contrast, \appname achieves the highest downstream resolution rate while retaining nearly twice as many resolved trajectories as SPICE, demonstrating a better balance between filtering data and preserving useful training trajectories.

\answerRQ{
Filtering SWE-rebench with \appname enables Qwen3.5-9B-Base to achieve a 66.2\% resolution rate on SWE-bench Verified, outperforming the best-performing baseline by 14.5\% while reducing the cost per resolved trajectory to \$0.46.
}

\begin{figure*}[!t]
    \centering
    \includegraphics[width=\linewidth]{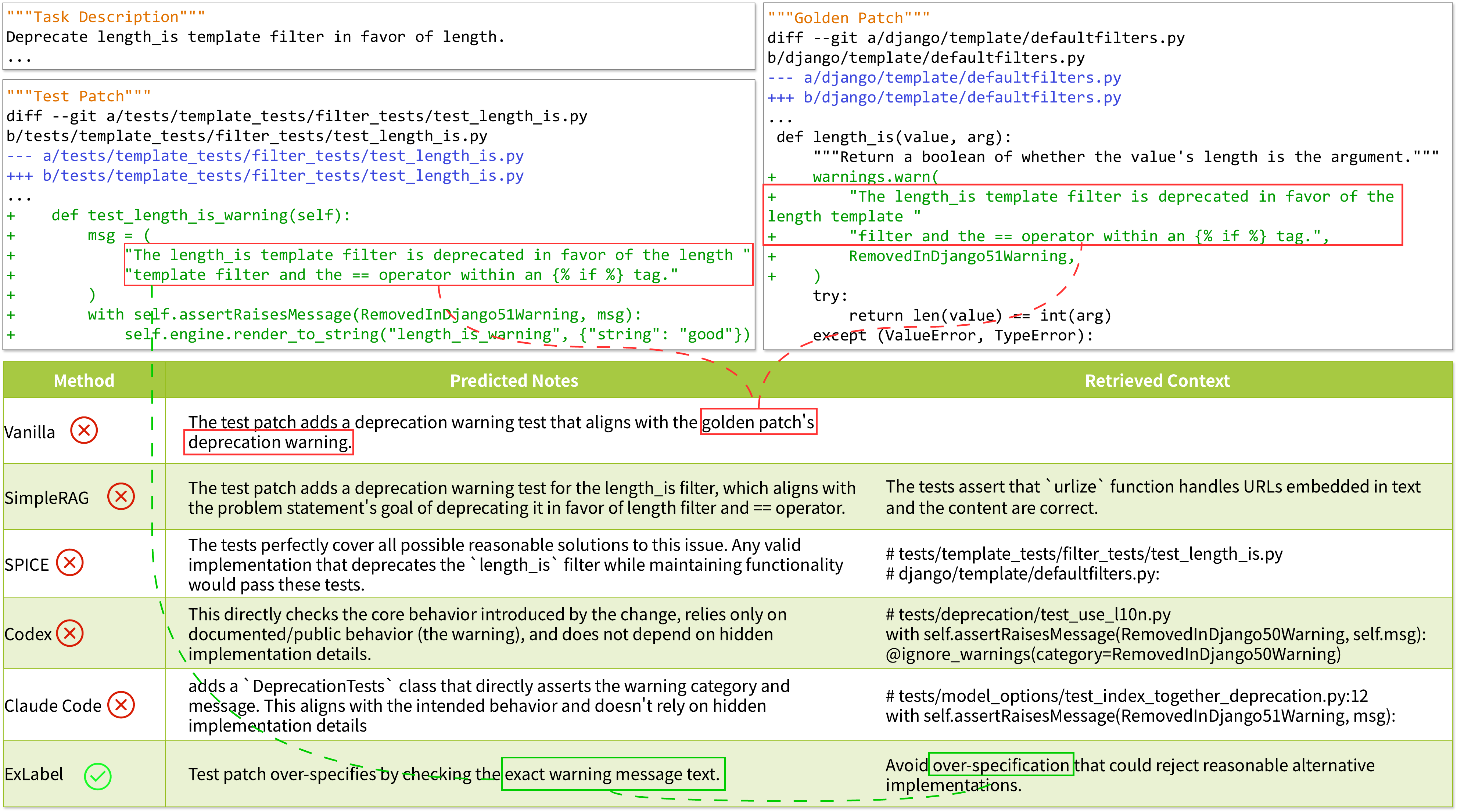}
    \caption{A test coverage assessment case of \texttt{django-15869} to reveal the reason for \appname's better performance.}
    \label{fig: effective_case}
\end{figure*}

\subsection{RQ4: Case Study}\label{rq4}

To understand how the experience of \appname helps it outperform the baselines, we manually inspect representative cases in RQ1 and find that the experience provides annotation-specific knowledge that additional context alone cannot offer.
For example, in the test coverage case shown in Figure~\ref{fig: effective_case}, the task description only asks to deprecate the \texttt{length\_is} template filter, while the test patch uses \texttt{assertRaisesMessage} to check the exact warning message implemented in the golden patch.
Therefore, a valid solution that raises the same warning with a different message would fail the test, making the test suite inadequate.
Vanilla notices that the test patch aligns with the warning in the golden patch but treats this alignment as evidence of adequacy, overlooking that such alignment is exactly the source of over-specification.
SimpleRAG retrieves a historical instance about the \texttt{urlize} function, whose annotation concerns URL handling and provides no guidance on over-specified tests.
SPICE locates the relevant test file and source file but still concludes that the tests cover all reasonable solutions, indicating that access to the relevant code alone is insufficient to judge whether the tests are over-specified.
Codex and Claude Code explore the repository and find other deprecation tests that also assert exact warning messages, and thus regard the message as intended behavior rather than an implementation detail.
However, such repository conventions only reflect how developers write tests, rather than whether the tests accept all valid solutions to the task.
In contrast, \appname retrieves the experience ``Avoid over-specification that could reject reasonable alternative implementations'' distilled from historical annotations, which directs the LLM to examine whether the test constrains details beyond the task description and thus identify the check on the exact warning message.
This case illustrates that \appname's experience provides more effective judgment criteria that may not be readily available from the instance itself, similar historical instances, or repository context.

\section{Discussion}

\subsection{Typical Annotation Errors in SWE Datasets}
\label{Dataset Re-annotation}

\begin{figure}[t!]
    \centering
    \includegraphics[width=\linewidth]{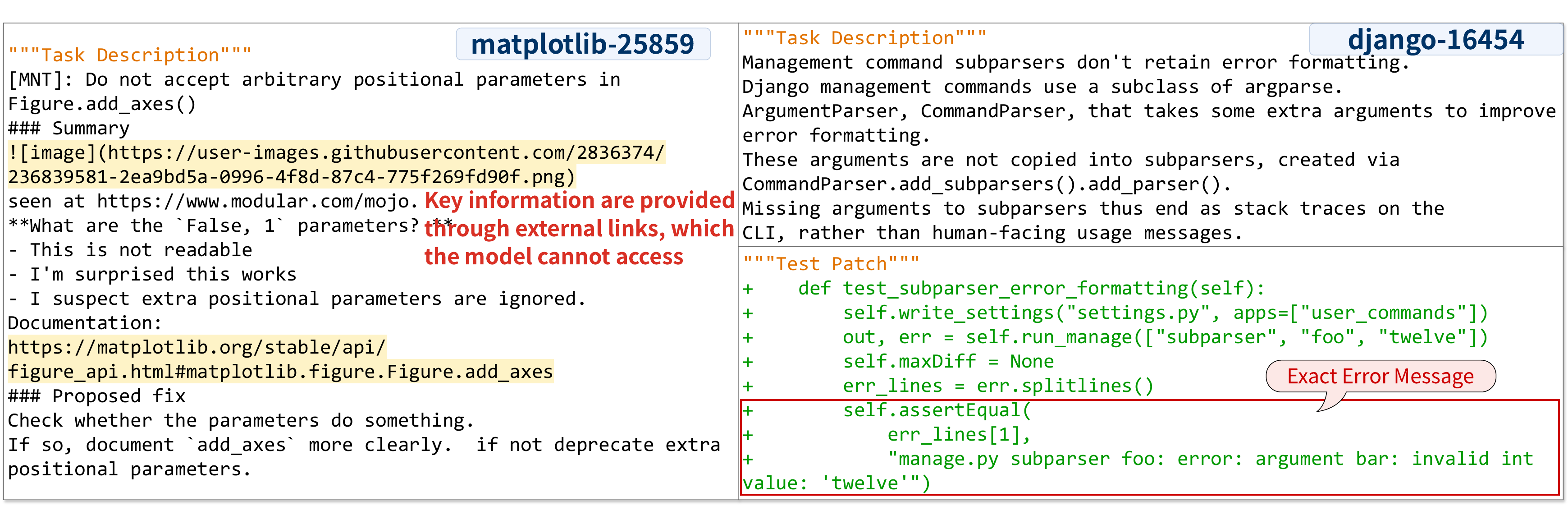}
    \caption{Examples of typical annotation error patterns in task clarity (left) and test coverage assessments (right).}
    \label{fig: inconsistent_cases}
\end{figure}

During the manual cleaning of \bench, we have observed the presence of incorrect annotations in the datasets for both task clarity and test coverage assessments, consistent with the observations of OpenAI’s annotation team~\cite{swebenchverified}, mainly manifested as follows.

For task clarity assessment, a typical annotation error occurs in instances whose task descriptions provide key information through external links~\cite{swebenchverified}.
Since LLMs are required not to access the internet when completing the task, information provided through external links is invalid~\cite{SWE-bench, choi2025contaminated, deng2024investigating}.
Therefore, such task descriptions should be labeled as underspecified.
Yet, they are incorrectly labeled as well-specified, which is inconsistent with the criteria for task clarity. 
For example, as shown in Figure~\ref{fig: inconsistent_cases} (left), the task description contains three external links that document code snippets relevant to the task and their corresponding documentation.
However, since the LLM cannot access these external links, the information within them cannot be obtained, resulting in missing key information in the task description. 
Without access to the code snippets corresponding to the images, it is nearly impossible for the LLM to understand what the parameters ``False, 1'' refer to, thereby rendering the task unsolvable.

For test coverage assessment, we further filter another type of annotation error, \textit{i.e.,} instances labeled as adequate even though the task description does not specify the exact error message assumed by the test patch.
Specifically, the golden patch implements the logic of raising an exception for certain inputs, whereas the corresponding test patch performs assertions by matching the error message as a string, making the test suite overly specific.
For example, as shown in Figure~\ref{fig: inconsistent_cases} (right), the test patch checks for specific error messages such as ``manage.py subparser foo: error: argument bar: invalid int value: \textquotesingle twelve\textquotesingle''. 
But this error message is not provided in the task description, which causes an LLM that generates a similar error message to be judged as failing the task, even when it has in fact solved it, simply because it cannot pass the test.
This is a typical instance of an inadequate test suite, yet it is incorrectly annotated as adequate.

By identifying and filtering the annotation errors of these two widely acknowledged patterns~\cite{swebenchverified}, we construct \bench as a high-quality benchmark for evaluating automated annotation methods, thereby ensuring the validity of the evaluation.

\subsection{Threats to Validity}

\textbf{Internal Validity.}
The datasets used in our experiments were manually annotated by the annotators of SWE-bench Verified and Multi-SWE-bench.
However, manual annotation may still contain errors.
To mitigate this threat, we retain only the instances on which all annotators reach complete agreement, and two of the authors further review all instances following the annotation criteria released by OpenAI~\cite{swebenchverified} to filter out erroneous annotations, achieving a Cohen's kappa of 0.928.

\noindent\textbf{External Validity.}
Our evaluation focuses on issue resolution, as existing SWE benchmarks are predominantly constructed around this task setting.
Nevertheless, issue resolution encompasses diverse SE scenarios, including bug fixing, feature implementation, and feature optimization~\cite{multiswebench}, allowing \bench to evaluate \appname across these diverse scenarios.
Moreover, \appname is not inherently tied to issue resolution, as its general interfaces and easily adaptable prompt templates facilitate its extension to other task settings.


\section{Related Work}

\subsection{SWE-bench-style Datasets}

With the rapid advancement of LLMs for SE tasks, increasing research attention has been devoted to constructing datasets for training and evaluating LLMs on diverse tasks (\textit{e.g.,} bug fixing and feature requests) in real-world software repositories.
A prominent family of such datasets is SWE-bench-style datasets.
Unlike traditional static code collections~\cite{liu2023your, jain2024livecodebench, zhuo2024bigcodebench}, SWE-bench-style datasets provide each instance with an executable environment, allowing LLM-based agents to interact with the repository, resolve tasks based on natural-language descriptions, and validate their solutions through execution.
Existing SWE-bench-style datasets can broadly be divided into benchmark and training datasets.

Benchmark datasets are typically collected from real-world repositories and equipped with executable environments.
For example, SWE-bench~\cite{SWE-bench} collects 2,294 executable instances from 12 real-world software repositories to evaluate LLMs on issue resolution.
Subsequent studies extend SWE-bench to multilingual~\cite{multiswebench, liveswebench} and multimodal~\cite{swebenchmultimodal} settings.
SWE-Perf~\cite{Swe-perf} and FEA-bench~\cite{fea-bench} further target performance optimization and feature addition, respectively, broadening the task scope while retaining the basic SWE-bench setting.
Training datasets additionally employ automated or synthetic construction to increase data scale.
For example, SWE-Smith~\cite{Swe-smith} introduces a reverse synthesis approach that starts from runnable code and automatically injects defects to generate new tasks.
Similarly, SWE-Synth~\cite{SWE-Synth} uses LLMs to simulate the debugging process, generating code fixes, test cases, and repair trajectories.
BugPilot~\cite{Bugpilot} synthesizes data by guiding coding agents to introduce new features into codebases and leveraging unintended test failures arising during this process to construct new tasks.

Despite substantial efforts devoted to constructing SWE-bench-style datasets~\cite{liveswebench, Swe-mirror, SWE-Lego, Swe-smith}, quality issues such as underspecified task descriptions and inadequate test suites have been observed.
For example, even SWE-bench Verified, which undergoes manual verification~\cite{swebenchverified}, has been reported to contain instances with insufficient test coverage~\cite{wang2025solved, yu2025utboost}.
Automated and synthetic construction may further introduce quality risks as datasets scale~\cite{Swe-smith, Swe-mirror}.
These concerns motivate techniques for systematically assessing task quality before SWE datasets are used for evaluation or training.

\subsection{Quality Assessment of SWE Datasets}
Before task clarity and test coverage were explicitly introduced as criteria for verifying SWE-bench instances~\cite{swebenchverified}, prior work had explored automated quality assessment for various software artifacts.
For example, Sheikhaei \textit{et al.}~\cite{sheikhaei2024empirical} investigated automated labeling of self-admitted technical debt (SATD) in code comments, comparing in-context learning with fine-tuned base models and showing the potential of LLMs for automated software-artifact assessment.
For SWE-bench-style datasets, however, task quality assessment remains challenging because determining whether a task is sufficiently specified and adequately tested requires contextual judgment over the task description and associated patches~\cite{spice}.
In practice, existing SWE datasets largely rely on human annotation and verification for task quality assessment~\cite{swebenchverified, nocode-bench}.
For example, OpenAI~\cite{swebenchverified} recruited 93 professional Python developers to independently annotate 1,699 SWE-bench instances, resulting in the SWE-bench Verified subset.
Multi-SWE-bench~\cite{multiswebench} and Doc2Feat-bench~\cite{nocode-bench} similarly employ manual annotation based on the criteria introduced by SWE-bench Verified.

Moreover, SPICE~\cite{spice} makes an initial attempt at automated annotation for SWE-bench-style datasets. 
It employs separate pipelines for task clarity and test coverage assessment: the former uses manually summarized patterns of underspecified task descriptions, while the latter leverages agent-based repository exploration to assess test adequacy.
However, such approaches assess individual instances using predefined criteria or instance-specific context, without explicitly accumulating and reusing knowledge learned from assessments across instances.
In contrast, \appname introduces a self-evolving experience bank that continuously extracts generalizable judgment criteria from historical experience and retrieves relevant experience to guide new instances.
This enables task quality assessment to benefit from cross-instance knowledge rather than treating each instance independently.

\section{Conclusion}

In this work, we propose \appname, an LLM-based, self-evolving, experience-guided assistant to assess task quality in SWE datasets.
\appname extracts common patterns and typical error patterns from historically annotated data through a self-evolving loop to form experience, thereby overcoming the limitations on cross-instance information utilization imposed by the stateless nature of LLMs.
We evaluate task quality from two perspectives: task clarity and test coverage, which concern solvability and verifiability, respectively.
To support the evaluation, we clean the human annotations of SWE-bench and Multi-SWE-bench and construct \bench, a benchmark spanning eight programming languages.
The experimental results on \bench show that \appname achieves the best F1 in all settings, with relative improvements in average F1 of 44.2\%/49.9\% and 24.0\%/18.4\% over the best-performing baseline for task clarity and test coverage assessment in \benchpyml, respectively.
Moreover, filtering the training data with \appname improves the fine-tuned coding model's resolution rate on SWE-bench Verified by 14.5\% over the best-performing baseline.


\section*{Data Availability}
\phantomsection\label{rp}
The replication package, including the \bench benchmark, the detailed prompt templates, and the appendix, is available at \url{https://github.com/ZJU-CTAG/ExLabel}.


\bibliographystyle{plainnat}
\bibliography{references}


\newpage
\appendix
\section*{Appendix}

\section{Detailed Results of RQ1}
\label{app: detailed}

\begin{table*}[h!]
\centering
\caption{Detailed results of RQ1, where P and R denote precision and recall, respectively, Python and Multilingual denote \benchpy and \benchml, respectively, and all metrics are reported in percentage.}
\label{table: full metrics}
\setlength{\tabcolsep}{4pt}
\begin{adjustbox}{width=\textwidth}
\begin{tabular}{cclcccc|cccc|cccc}
\toprule
& & \multirow{2}{*}{\textbf{Method}}
& \multicolumn{4}{c}{\textbf{Qwen3.8-27B}} & \multicolumn{4}{c}{\textbf{DeepSeek-V4-Flash}} & \multicolumn{4}{c}{\textbf{GPT-5.6 Luna}} \\
\cmidrule(lr){4-7} \cmidrule(lr){8-11} \cmidrule(lr){12-15}
& & & F1$\uparrow$ & P$\uparrow$ & R$\uparrow$ & FPR$\downarrow$ & F1$\uparrow$ & P$\uparrow$ & R$\uparrow$ & FPR$\downarrow$ & F1$\uparrow$ & P$\uparrow$ & R$\uparrow$ & FPR$\downarrow$ \\
\midrule
 &  & Vanilla & 50.0 & \textbf{100.0} & 33.3 & \textbf{0.0} & 24.7 & 16.4 & 50.0 & 16.3 & 45.5 & 38.5 & 55.6 & 5.7 \\
 &  & SimpleRAG & 43.5 & \textbf{100.0} & 27.8 & \textbf{0.0} & 19.6 & 11.3 & \underline{72.2} & 36.2 & 48.7 & 47.4 & 50.0 & \underline{3.6} \\
 &  & SPICE & \underline{52.8} & 40.0 & \textbf{77.8} & 7.4 & \underline{39.6} & 24.7 & \textbf{100.0} & 19.5 & 30.8 & 18.2 & \textbf{100.0} & 28.7 \\
 &  & Codex & 20.0 & \textbf{100.0} & 11.1 & \textbf{0.0} & 30.3 & \underline{33.3} & 27.8 & \underline{3.6} & \underline{52.6} & \underline{50.0} & 55.6 & \underline{3.6} \\
 &  & Claude Code & 20.0 & \textbf{100.0} & 11.1 & \textbf{0.0} & 37.2 & 32.0 & 44.4 & 6.0 & 47.6 & 41.7 & 55.6 & 5.0 \\
\rowcolor{paleGreen}
\cellcolor{white} & \cellcolor{white}\multirow{-6}{*}{\rotatebox[origin=c]{90}{Python}} & \appnamebold & \textbf{64.7} & \underline{68.8} & \underline{61.1} & \underline{1.8} & \textbf{51.9} & \textbf{77.8} & 38.9 & \textbf{0.7} & \textbf{61.1} & \textbf{61.1} & \underline{61.1} & \textbf{2.5} \\
\cmidrule(lr){2-15}
 &  & Vanilla & 26.7 & 23.5 & 30.8 & 4.8 & 15.0 & 8.7 & 53.9 & 26.9 & 17.3 & 10.5 & \underline{50.0} & 20.3 \\
 &  & SimpleRAG & \underline{26.9} & 26.9 & 26.9 & 3.5 & 12.2 & 6.6 & \textbf{76.9} & 51.7 & \underline{20.8} & \underline{14.3} & 38.5 & \underline{11.0} \\
 &  & SPICE & 13.9 & 8.3 & \textbf{42.3} & 22.1 & 15.4 & 8.7 & \underline{65.4} & 32.5 & 13.2 & 7.3 & \textbf{73.1} & 44.4 \\
 &  & Codex & 17.1 & 33.3 & 11.5 & \underline{1.1} & \underline{16.6} & 10.1 & 46.2 & 19.6 & 14.5 & 9.2 & 34.6 & 16.3 \\
 &  & Claude Code & 24.2 & \textbf{57.1} & 15.4 & \textbf{0.6} & 16.2 & \underline{11.0} & 30.8 & \underline{11.9} & 13.3 & 8.0 & 38.5 & 21.0 \\
\rowcolor{paleGreen}
\cellcolor{white}\multirow{-12}{*}{\rotatebox[origin=c]{90}{Task Clarity}} & \cellcolor{white}\multirow{-6}{*}{\rotatebox[origin=c]{90}{Multilingual}} & \appnamebold & \textbf{37.5} & \underline{40.9} & \underline{34.6} & 2.4 & \textbf{25.4} & \textbf{20.0} & 34.6 & \textbf{6.6} & \textbf{26.9} & \textbf{22.0} & 34.6 & \textbf{5.9} \\
\midrule
 &  & Vanilla & 10.0 & 75.0 & 5.4 & \underline{0.9} & 13.3 & \textbf{100.0} & 7.1 & \textbf{0.0} & 12.7 & 57.1 & 7.1 & \textbf{2.8} \\
 &  & SimpleRAG & 66.7 & \textbf{93.6} & 51.8 & 1.8 & 31.9 & \underline{84.6} & 19.6 & \underline{1.8} & 20.9 & 63.6 & 12.5 & \underline{3.7} \\
 &  & SPICE & \underline{71.3} & 63.0 & \textbf{82.1} & 24.8 & \underline{37.5} & 62.5 & \underline{26.8} & 8.3 & \underline{75.8} & 65.8 & \underline{89.3} & 23.9 \\
 &  & Codex & 10.0 & 75.0 & 5.4 & \underline{0.9} & 16.9 & 40.0 & 10.7 & 8.3 & 43.6 & \textbf{77.3} & 30.4 & 4.6 \\
 &  & Claude Code & 0.0 & 0.0 & 0.0 & \textbf{0.0} & 9.2 & 33.3 & 5.4 & 5.5 & 30.6 & 68.8 & 19.6 & 4.6 \\
\rowcolor{paleGreen}
\cellcolor{white} & \cellcolor{white}\multirow{-6}{*}{\rotatebox[origin=c]{90}{Python}} & \appnamebold & \textbf{76.5} & \underline{84.8} & \underline{69.6} & 6.4 & \textbf{73.0} & 71.2 & \textbf{75.0} & 15.6 & \textbf{79.4} & \underline{69.3} & \textbf{92.9} & 21.1 \\
\cmidrule(lr){2-15}
 &  & Vanilla & 24.4 & 50.0 & 16.1 & \underline{3.5} & \underline{33.7} & \textbf{43.6} & \underline{27.4} & \underline{7.7} & 20.3 & \underline{47.1} & 12.9 & \textbf{3.2} \\
 &  & SimpleRAG & 26.2 & 26.7 & 25.8 & 15.4 & 13.3 & 38.5 & 8.1 & \textbf{2.8} & 36.0 & \textbf{47.4} & 29.0 & \underline{7.0} \\
 &  & SPICE & \underline{31.7} & 22.9 & \textbf{51.6} & 37.8 & 27.0 & 24.1 & \textbf{30.7} & 21.0 & 35.7 & 24.0 & \textbf{69.4} & 47.6 \\
 &  & Codex & 14.1 & \underline{55.6} & 8.1 & \textbf{1.4} & 19.1 & 23.3 & 16.1 & 11.5 & \underline{38.9} & 45.7 & 33.9 & 8.7 \\
 &  & Claude Code & 16.7 & \textbf{60.0} & 9.7 & \textbf{1.4} & 8.9 & 14.3 & 6.5 & 8.4 & 35.2 & 41.3 & 30.7 & 9.4 \\
\rowcolor{paleGreen}
\cellcolor{white}\multirow{-12}{*}{\rotatebox[origin=c]{90}{Test Coverage}} & \cellcolor{white}\multirow{-6}{*}{\rotatebox[origin=c]{90}{Multilingual}} & \appnamebold & \textbf{36.9} & 46.3 & \underline{30.7} & 7.7 & \textbf{34.9} & \underline{40.4} & \textbf{30.7} & 9.8 & \textbf{40.0} & 30.1 & \underline{59.7} & 30.1 \\
\bottomrule
\end{tabular}
\end{adjustbox}
\end{table*}

In addition to F1 and FPR, Table~\ref{table: full metrics} reports the precision and recall of all techniques in RQ1.

Table~\ref{table: full metrics} further reveals different precision-recall trade-offs among the techniques.
Vanilla, SimpleRAG, Codex, and Claude Code tend to be conservative, and their average recall is only 22.3\%--36.6\% across the 12 combinations of tasks, subsets, and backbone models.
In all 17 cases where one of these baselines achieves a precision no lower than that of \appname, its recall is lower than that of \appname, such as the precision of 100.0\% with a recall of only 11.1\%--33.3\% on task clarity assessment in \benchpy with Qwen3.8-27B.
Such conservative predictions leave most low-quality instances in the dataset.
In contrast, SPICE achieves the highest recall in 9 of the 12 combinations with an average recall of 67.4\%, but its average precision is only 30.8\%, which is the lowest among all techniques.
For example, SPICE achieves a recall of 73.1\% but a precision of only 7.3\% on task clarity assessment in \benchml with GPT-5.6 Luna, indicating that most of the instances it flags are valid ones.
\appname strikes a better balance between the two, achieving the highest average precision of 52.7\% and an average recall of 52.0\%, which is second only to SPICE.
The average gap between the precision and recall of \appname is 14.8 percentage points, compared with 29.4--42.5 percentage points for the baselines.
Meanwhile, the trade-off of \appname varies with the backbone model.
With GPT-5.6 Luna, \appname favors recall on test coverage assessment, achieving recalls of 92.9\% and 59.7\% with precisions of 69.3\% and 30.1\% in \benchpy and \benchml, respectively, which explains its relatively high FPR in RQ1.
Overall, this balance enables \appname to remove more low-quality instances than the conservative techniques while discarding fewer valid instances than SPICE, which is consistent with its downstream benefits in RQ3.

\section{Impact of Historical Data Size}
\label{app: scaling}

\begin{table}[h!]
\centering
\caption{F1 of \appname with GPT-5.6 Luna using different proportions of the historical data for experience accumulation.}
\label{table: scaling}
\setlength{\tabcolsep}{6pt}
\begin{tabular}{llcccccc}
\toprule
\textbf{Task} & \textbf{Subset} & 0\% & 20\% & 40\% & 60\% & 80\% & 100\% \\
\midrule
\multirow{2}{*}{Task Clarity} & Python & 45.5 & \textbf{66.7} & 64.7 & 56.3 & 57.1 & 61.1 \\
 & Multilingual & 17.3 & 20.0 & 23.2 & 15.9 & 22.0 & \textbf{26.9} \\
\midrule
\multirow{2}{*}{Test Coverage} & Python & 12.7 & \textbf{80.0} & 77.5 & 77.9 & 79.7 & 79.4 \\
 & Multilingual & 20.3 & 38.9 & 37.3 & 39.1 & 39.1 & \textbf{40.0} \\
\bottomrule
\end{tabular}
\end{table}

To investigate how much annotated data is required for the experience bank to become effective, we accumulate experience on 20\%, 40\%, 60\%, and 80\% of the historical data for each task and evaluate \appname with GPT-5.6 Luna on the same test data as RQ1.
As shown in Table~\ref{table: scaling}, accumulating experience on 20\% of the historical data, \textit{i.e.,} about 400 annotated instances, improves F1 by 2.7--67.3 percentage points over w/o Experience Bank in all four combinations of tasks and subsets.
On task clarity assessment, the performance of \appname tends to stabilize early, as F1 shows no clear increase beyond 20\% of the historical data.
This is likely because underspecified instances account for only 4.9\% of the historical data, and their typical causes, such as missing key information in the task description, mainly concern the description itself, so the experience bank can capture them from a small amount of data while additional data may contribute limited new experience.
On test coverage assessment, F1 shows an increasing trend from 40\% to 100\% of the historical data, rising from 77.5\% to 79.4\% and from 37.3\% to 40.0\% in \benchpy and \benchml, respectively.
This is likely because inadequate instances account for 32.1\% of the historical data, and their causes, such as test patches relying on specific error messages or implementation details, depend on the golden patch, the test patch, and the evolving state of the codebase, which tend to vary across repositories, so additional data may still contribute new experience.

\begin{figure*}[h]
    \centering
    \includegraphics[width=1\linewidth]{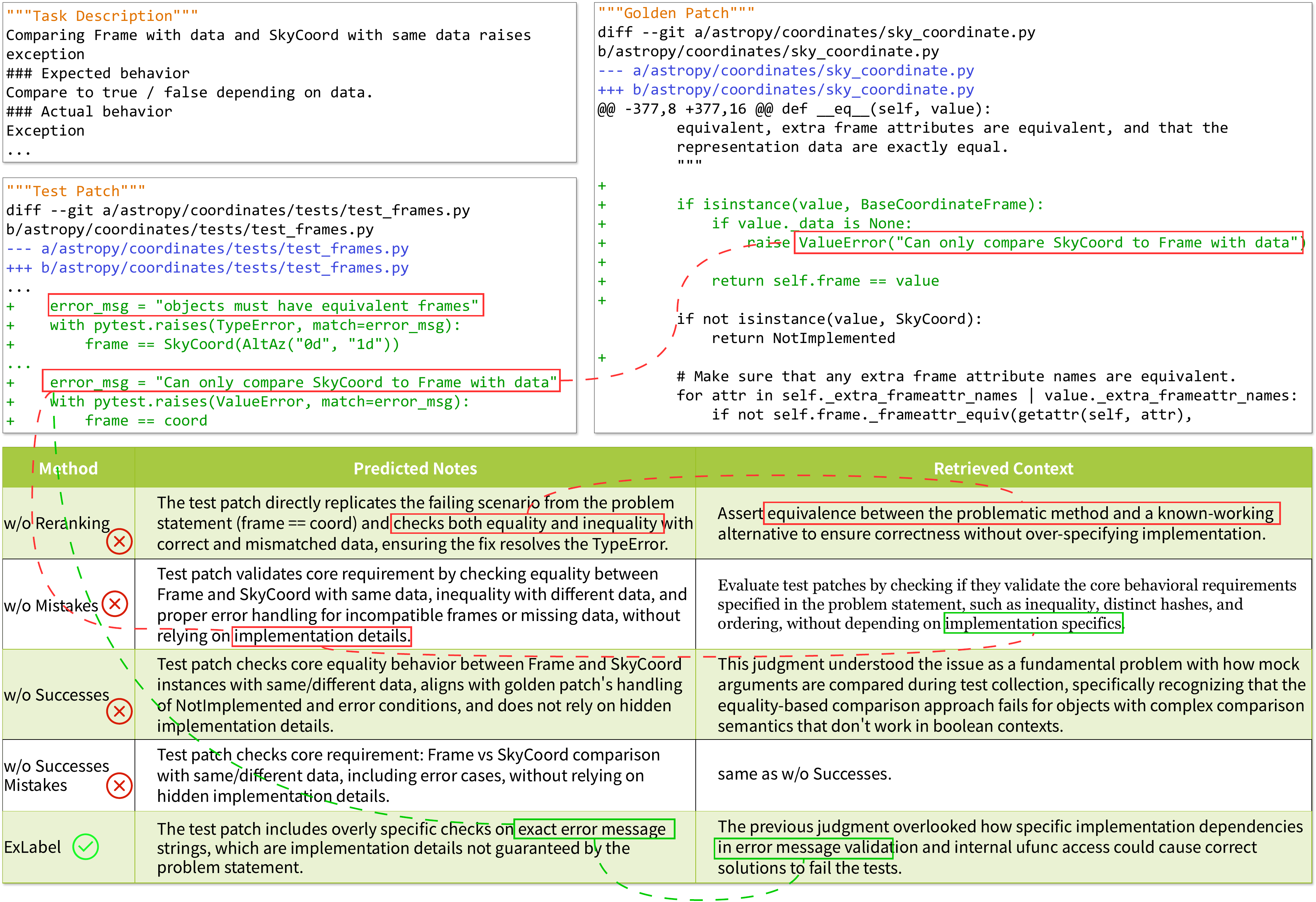}
    \caption{A test coverage assessment case of \texttt{astropy-13477} to reveal the effectiveness of each component in \appname.}
    \label{fig: ablation_case}
\end{figure*}
\section{Training Recipe of RQ3}
\label{app: training}
In RQ3, we perform full-parameter fine-tuning on Qwen3.5-9B-Base~\cite{qwen35} using the LLaMA-Factory~\cite{zheng2024llamafactory} framework on 4 H100 GPUs.
Following prior work~\cite{swegym, Swe-smith, SWE-Lego}, we set the batch size to 8, the learning rate to $5\times10^{-5}$, the maximum number of epochs to 3, and the maximum context length to 131,072 tokens, use a cosine learning-rate scheduler, and set the warmup ratio to 0.1.

\section{Case Study for Ablation}
\label{app: ablation case}
We present an additional representative case for test coverage assessment in RQ2, as shown in Figure~\ref{fig: ablation_case}.  
In this case, the test patch depends on a specific error string implemented in the golden patch, namely ``Can only compare SkyCoord to Frame with data'', which is not explicitly stated in the task description.  
For \textbf{w/o Reranking}, the retrieved experience emphasizes the equivalence between the problematic method and a known-working alternative.  
This shifts the LLM’s focus away from dependence on a specific implementation toward verifying behavioral consistency and semantic correctness.  
For \textbf{w/o Field (Mistakes)}, the retrieved experience focuses on the core behavior requirement.  
Although the experience mentions implementation details, it lacks further elaboration. 
As a result, the LLM directly concludes that there is no dependency on implementation details after checking the core requirement.  
For \textbf{w/o Field (Successes)} and \textbf{w/o Field (Successes, Mistakes)}, both settings retrieve identical experience content, which is an understanding of an unrelated issue about how mock arguments are compared during test collection.
This content lacks direct guidance for the current instance.
Consequently, the LLM fails to identify the dependency on a specific error message and instead makes a judgment solely based on core behavior.  
In contrast, \appname retrieves experience whose \textbf{Mistakes} field explains that a previous judgment overlooked how implementation dependencies in error message validation could cause correct solutions to fail the tests.
This experience enables the model to identify the hidden dependency and generate the correct prediction successfully.

\end{document}